\documentclass[preprint,12pt]{elsarticle}

\usepackage{amssymb}
\usepackage{amsmath}
\usepackage{epsfig}
\usepackage{graphicx}
\usepackage{dcolumn}
\usepackage{bm}
\usepackage{mathrsfs}
\usepackage{amssymb}
\usepackage{float}
\usepackage{cancel}
\usepackage[colorlinks=true, citecolor=red]{hyperref}
\usepackage{enumitem}

\journal{Optics Communications}

\begin{document}

\begin{frontmatter}



\title{Continuous Recovery of Phase from Single Interferogram}


\author[label1]{V. Berejnov} 
\ead{berejnov@gmail.com} 

\author[label2]{B.Y. Rubinstein} 
\ead{bru@stowers.org} 

\affiliation[label1]{
            addressline={6575 St. Charles Pl}, 
            city={Burnaby},
            postcode={V5H3W1}, 
            state={BC},
            country={Canada}}
\affiliation[label2]{
            addressline={Stowers Institute for Medical Research 1000 50th St.}, 
            city={Kansas City},
            postcode={64110}, 
            state={MO},
            country={USA}}

\begin{abstract}
A new method for phase recovery from a single two-beam interferogram is presented. 
Conventional approaches, relying on trigonometric inversion followed by phase unfolding and unwrapping, are hindered by discontinuities typically addressed through intricate algorithms. Our method bypasses the unfolding and unwrapping, instead formulating a first-order differential equation directly relating the phase to the interferogram. Integration of this equation enables continuous retrieval of phase  along any  straight path. Representing a new class of analytical tools for single-interferogram phase retrieval, this approach is derived from first principles and accommodates both Newton-type and Fizeau-type interferograms. Its performance is demonstrated on multiple idealized synthetic interferograms of increasing complexity, validating against the known seed phase.
\end{abstract}




\end{frontmatter}



\section{Introduction}

Interferometry has long been a direct characterization experimental technique, providing 
an access to the shape of light-reflecting surfaces and the optical density of light-refracting 
media. The range of characterized objects may include mirrors, lenses \cite{Malacara2005}, 
thin films and layered materials \cite{Gokhale2004}, as well as more complex systems such as inhomogeneous 
gas or liquid flows. In these cases, the measured quantity of light—a phase, may correspond to surface 
topography \cite{DellAversana1997}, mass density \cite{Mialdun2011}, or temperature 
distributions \cite{Bratukhin2005}. The interferometric measurements can be conducted 
under steady-state conditions or with short time exposures for capturing the dynamic processes.

The simplest and fastest interferometric measurement is a single interferogram represented by an image formed under monochromatic illumination by two or more beams that records in one exposure their phase difference in terms of interference fringes. These fringes vary in intensity between dark (minima) and bright (maxima), often forming complex patterns across the interferogram image plane. Their shape and 
intensity encode the spatial variation of the pointwise phase difference between the interfering light beams, which is a main output of interferometry. 
The spatial phase variation contained in the interferogram can be converted 
into meaningful physical quantities, such as surface topography, thickness maps, or flow field characteristics.

The central problem is an extraction of the spatial phase variation from a single interferogram. 
Determining the spatial phase difference \footnote{Hereafter, we distinguish two uses of “phase difference”: (i) in a single point (pointwise), denoting the phase offset between two interfering beams; (ii) between two points, denoting spatial variation of the phase profile. The meaning depends on context.}
between maxima (or minima) of two adjacent fringes is straightforward, it equals $2\pi$, a full phase cycle. In contrast, evaluating the phase difference between two {\it arbitrary} points in an interferogram is more demanding. The challenge lies in counting the number of full $2\pi$ cycles along the path connecting the points, identifying the fractional position of each point within its fringe, and tracking the direction of phase change (increasing or decreasing) along that path.  

\begin{figure}[!htbp]
\centering
\includegraphics[width=0.8\linewidth]{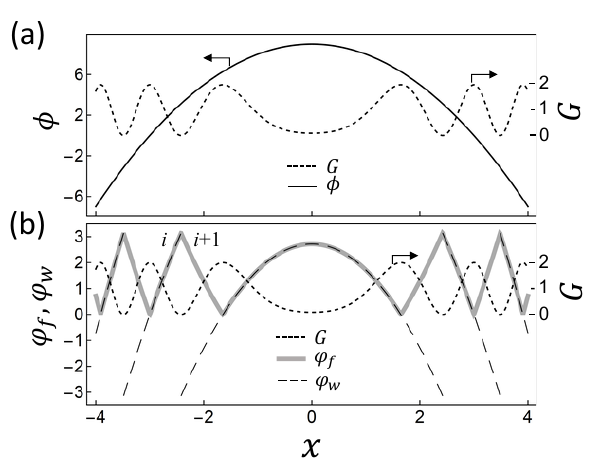}
\caption{
Illustration of one-dimensional phase recovery.  
(a) The initial continuous phase $\phi$ and its corresponding interferogram $G$.  
(b) Folded phase $\varphi_f$, obtained as $\arccos(G-1)$, and wrapped phase $\varphi_w$ obtained by unfolding each pair $\{\varphi_{f,i},\varphi_{f,i+1}\}$.  
The wrapped phase represents a sequence of discontinuous pieces of the initial phase $\phi$ mutually shifted by $2\pi$.}
\label{fig0}
\end{figure}

This is illustrated by the idealized one-dimensional model of a single interferogram \cite{Judge1994}, where all variables are related to the point $x$ on the interferogram 
\begin{equation*}
G = A + B \cos\phi,
\end{equation*}
here, $G$ represents the recorded intensity of the interferogram, coefficients $A$ and $B$ depend on the 
amplitudes of the reference and object beams, as well as the detector response and $\phi$ is phase, Fig.~\ref{fig0}(a). 

When the intergerogram $G$ is given,  
phase recovery traditionally starts with the trigonometric function inversion.  
Even in the simplest case, where $A=B=1$, such inversion yields not the continuous phase $\phi(x)$, but its folded version:
\begin{equation*}
\varphi_f(x)=\arccos(G(x)-1).
\end{equation*}
The phase $\varphi_f(x)$  is represented by an array of 
segments $\{\varphi_{f,i}(x)\}$ sequentially \textit{folded} into the principal range $[0,\pi]$ of the arccosine function~\cite{Okada2007}, Fig.~\ref{fig0}(b). Within each fully developed $G$-fringe, corresponding to a phase interval of $[0,2\pi]$, this folding produces two neighbouring sub-segments $\varphi_{f,i}$ and $\varphi_{f,i+1}$, each 
restricted to $[0,\pi]$. Their slopes are inverted with respect to each other, giving the characteristic sawtooth structure, Fig.~\ref{fig0}(b).

This folded phase needs to be unfolded \cite{Okada2007} transforming the entire $\varphi_f(x)$ by mirror reflecting one sub-segment from each pair $\varphi_{f,i}$ and $\varphi_{f,i+1}$. Such unfolding is a heuristic procedure requiring in general the knowledge of the phase derivative sign~\cite{Okada2007}. 
The resulting array $\varphi_w(x)=\{\varphi_{w,j}(x)\}$ represents a sequence of \textit{wrapped} phase pieces, Fig.~\ref{fig0}(b). 

While the wrapped phase is globally discontinuous, 
each segment $\varphi_{w,j}(x)$ remains continuous and 
correctly reproduces the spatial profile of the original phase $\phi(x)$ within the interval $[0,2\pi]$, Figs.~\ref{fig0}(a,b). To recover the 
complete phase underlying the interferogram $G$, these wrapped segments must be 
reassembled into a single continuous profile.

Converting $\varphi_{w,j}(x)$ into a continuous 
phase requires the use of a phase \textit{unwrapping algorithm}~\cite{Judge1994}. 
Numerous variations of unwrapping methods exist depending on the specific application, and can be found elsewhere \cite{GhigliaPritt1998,Baldi2001}. These algorithms
require different degree of heuristic assumptions, some could be applied 
to a single interferogram, while others require a set of interferograms with a known phase change. 

The illustrated phase transformation underscores 
an important methodological point: following the initial trigonometric inversion, 
phase recovery is necessarily represented by a \textit{discrete} mathematical 
framework relying on algorithms, performing phase unfolding and unwrapping.

It is striking that, even when both the reference and object beams are continuous 
and yield a continuous single interferogram, as in Fig.~\ref{fig0}(a), the underlying 
 phase cannot be {\it recovered continuously}.

This limitation has led to the widespread use of phase recovery algorithms or 
multiple-interferogram recording—methods so deeply embedded in practice that their necessity is rarely questioned. However, this dependence on trigonometric inversion is not fundamental, it arises from the conventional mathematical formulation rather than from intrinsic physical constraints.

In this paper, we introduce a new method for 
continuous retrieving a phase from a single interferogram. 
Unlike conventional approaches, this method eliminates trigonometric inversion 
by substituting it with a differential formulation, thereby bypassing the need for 
an heuristic algorithmic phase recovery tools and relying the same time on the single interfeogram only. 

The phase is recovered directly as a continuous function along any chosen straight path between two points on the interferogram. The method is analytical with a compact and self-contained mathematical formulation.  Being applied to  the specified interference conditions, it  produces a differential equation linking the phase and the interferogram function. Solution of  this equation, either analytically or numerically along the desired path, provides the continuous phase profile. We derive the method from first principles under simplified assumptions 
and demonstrate its  applicability across a range of representative interferograms using analytical and numerical examples. 

Our method operates on a single interferogram image. We describe the entire 
interferogram by an interferogram function $F(\bm r)$, where the argument $\bm{r}=(x, y)$ 
denotes the spatial coordinates in the interferogram plane, corresponding to pixel 
positions in a two-dimensional (2D)  Cartesian system $(x, y)$. 
The fringe system in the interferogram is 
governed by the unknown underlying phase distribution $\varphi (\bm r)$, which the 
method aims to recover from the known function $F(\bm r)$.

The method formulation depends on the conditions of the interferometric 
experiment—most notably, whether it involves two or multiple interfering beams, 
determining  specific forms of $F(\bm r)$, and whether the phase is spatially 
modulated, and/or  illumination background and noise exist. For instance, a \textit{linearly modulated} total phase results in a Fizeau-type interferogram fringe pattern, in contrast to an \textit{unmodulated} phase that  produces a zero-order fringe pattern (Newton-type fringes). While these conditions influence the mathematical structure of the method, its main principle remains applicable 
across all these cases.

For clarity and consistency across all presented  examples we focus 
on the case of two-beam interference and construct the corresponding function $F(\bm r)$, 
accordingly. We  begin with the examples having the interferograms representing 
the unmodulated phase $\varphi(\bm r)$, producing zero-order (Newton-type) fringes. This baseline 
consideration is both — simple and rather general, demonstrates the process of phase 
recovery. Then, introducing a spatial linear carrier we proceed to the modulated phase case to illustrate recovery  
of the total phase from the Fizeau-type interferogram.

All examples presented in this paper assume idealized illumination, resulting in a 
function $G$ with a \textit{uniform intensity envelope}. This means that any variation 
in fringe intensity arises only from the local (pointwise) phase difference of interfering beams. As a result, the intensity 
of fully developed fringes, those corresponding to a complete $2\pi$ phase difference, 
remains constant, while fringes associated with smaller phase differences appear with reduced intensity. 

The method can in principle be extended to the case of multiple-beam interference, as 
encountered in thin-film interference. It is also compatible with more realistic illumination, 
including the non-uniform background intensity and noise, resulting in a non-uniform $G$ 
in the recorded fringe pattern. These generalizations introduce additional complexities 
that are independent of the core method and will be briefly considered in the Discussion 
section. A comprehensive treatment of these extensions lies beyond the scope of this 
paper and will be published elsewhere.

The method is demonstrated using variety of synthetic interferograms with known 
phase profiles, referred to as the \textit{seed} phase. However, the phase is retrieved 
directly from the interferogram function $F(\bm r)$, without involving the seed phase 
in the reconstruction process. The seed phase is used only for benchmarking and validation of the results.

\section{Formulation}

Consider the interference of two monochromatic and coherent light waves each with smooth
complex amplitude $A_r$ and $A_s$, representing the 
reference and object (sample) beams, respectively. In practice, the reference 
beam is usually well defined and the object beam is a portion of the reference 
beam either reflected from or transmitted through the sample under study, 
experiencing no additional spatial modulation for now. At a given region in 
3D space, these two beams superimpose and their interference produces a 
spatially modulated intensity field $I$. A planar detector is placed in this 
region facing both beams to record the resulting interference pattern. The 
intensity recorded over the detector plane, expressed in terms of the detector 
optical density $G$, defines the interferogram.

Since the detector captures a 2D cross-section of the 3D intensity field, both 
$I$ and $G$ are defined within the coordinate system of the detector plane. 
Assuming a linear detector response, the optical density $G(\bm r)$ is 
proportional to the local intensity $I(\bm r)$, such that: $G(\bm r)=\eta I(\bm r)$, 
where $\eta$ is a detector efficiency factor, and $\bm r$ denotes the 2D spatial 
coordinate in the detector plane.

The local intensity is given by the square modulus of the superposed 
complex amplitudes: 
$I = |A_r + A_s|^2 = (A_r + A_s)(A_r + A_s)^*$,
where star denotes complex conjugation  and the complex amplitudes are 
expressed in exponential form as 
$A_r = a_r\, e^{i\varphi_r}$ and $A_s = a_s\, e^{i\varphi_s}$.  
Here, $a_r,\;a_s,\;\varphi_r$, and $\varphi_s$ 
represent the spatially 
varying amplitudes and phases of the reference 
and object beams at the detector plane at $\bm r$, respectively. Substituting the 
expressions for $A_r$ and $A_s$ 
into the intensity equation and simplifying via Euler's identitie yields
\begin{equation}
I = a_r^2 + a_s^2 + 2\, a_r a_s \cos\Delta\varphi,
\label{eq1}
\end{equation}
where $\Delta\varphi = \varphi_s - \varphi_r$ is the phase difference at each point 
of the detector plane. Therefore, the detector output (or interferogram optical density) 
takes the form
\begin{equation}
G(\bm r) = A(\bm r) + B(\bm r) \cos\Delta\varphi(\bm r),
\label{eq2}
\end{equation}
where the coefficient $A(\bm r) = \eta \bigl(a_r^2(\bm r) + a_s^2(\bm r)\bigr)$ 
represents the  background illumination, and 
the coefficient $B(\bm r) = 2 \eta\, a_r(\bm r) a_s(\bm r) $
modulates the fringe intensity independently of phase variations. Eq.~(\ref{eq2}) describes the gray-level image (interferogram) recorded by the detector. 

Eq.~(\ref{eq2}) is convenient for modeling the interferogram because the 
fringe pattern $G(\bm r)$ is a result of the functions $A(\bm r)$, $B(\bm r)$, 
and $\Delta \varphi(\bm r)$, which can be independently selected, bypassing 
the detailed consideration of the reference and object light waves. 

Collecting all terms related to the interferogram image we define an interferogram function for two-beam interference 
\begin{equation}
F(\bm r) =\frac{G(\bm r) - A(\bm r)}{B(\bm r)} 
\label{eq3}
\end{equation}
such that $F(\bm r) \in [-1, 1]$. Because the interferogram intensity is sensitive to the pointwise phase 
difference only, it is convenient to replace $\Delta \varphi$ by $\varphi$, assuming 
that $\varphi$ measures the spatial phase profile of the object beam relative to the reference 
beam. Then, from Eq.~(\ref{eq2}), the interferogram function satisfies
\begin{equation}
F(\bm r) = \cos \varphi(\bm r).
\label{eq4}
\end{equation}
Traditionally, solving Eq.~(\ref{eq4}) requires inverting the cosine function, 
which after a piecewise rearrangement (unfolding) yields a wrapped phase confined to 
the range $[-\pi, \pi]$. We   propose an
alternative strategy: differentiate Eq.~(\ref{eq4}) and use the resulting 
differential form to eliminate the trigonometric dependency.

Taking the derivative of Eq.~(\ref{eq4}) with respect to spatial position $\bm r$, 
we obtain for its two components  
\begin{equation}
F'_x =  -\sin \varphi(\bm r)\varphi'_x,\
F'_y =  -\sin \varphi(\bm r)\varphi'_y,
\label{eq5}
\end{equation}
where we introduce the notations 
$f_x' = \partial f(\bm r)/\partial x$ and $f_y' = \partial f(\bm r)/\partial y$ for the partial derivatives.  
Solving Eq.~(\ref{eq5}) for $\sin \varphi(\bm r)$ and using it with Eq.~(\ref{eq4}) in 
the Pythagorean identity $\sin^2 \alpha + \cos^2 \alpha = 1$, we eliminate the trigonometric 
functions from consideration and obtain two  first-order ordinary  differential equations (ODEs) for the phase
\begin{equation}
(\varphi'_x)^2 = \frac{(F'_x)^2}{1 - F^2}, \
(\varphi'_y)^2 = \frac{(F'_y)^2}{1 - F^2}.
\label{eq6}
\end{equation}
The \textit{phase-retrieving equations} ~(\ref{eq6}) form   the core of the proposed method, 
linking the interferogram function representing the fringe pattern and the smooth phase underlying 
the interferogram. Note, Eqs.~(\ref{eq6}) describe the phase $\varphi(\bm r)$ as a continuous 
function, recovered directly from the continuous interferogram function $F(\bm r)$, without the need for unfolding and unwrapping. 
Together, Eqs.~(\ref{eq3}) and~(\ref{eq6}) constitute a closed system: 
once the interferogram $G(\bm r)$ is transformed into $F(\bm r)$, the phase can in principle 
be recovered by solving Eqs.~(\ref{eq6}). Eq.~(\ref{eq4}), serving as a starting point of 
the method, will also be used to set a boundary condition; see details in the Solution section.

\section{Solution}

Consider the interferogram fringe pattern $G(x, y)$ in the   Cartesian coordinate system. The gray 
values $G(x, y)$ oscillate between minimum (black) and maximum (white) pixel intensities in a 
finite region within a boundary defined as follows: 
${\cal D} = \{ (x, y) : x_{\min} \leq x \leq x_{\max},\ y_{\min} \leq y \leq y_{\max} \}$. For the 
phase-retrieving task, we need to combine the interference function specified by the experiment, 
Eq.~(\ref{eq3}), and the phase-retrieving equations, Eqs.~(\ref{eq6}):
\begin{equation}
F(x, y) = \frac{G(x, y) - A(x, y)}{B(x, y)},
\label{eq7}
\end{equation}
\begin{equation}
\begin{aligned}
\big(\varphi_x'(x, y) \big)^2 = \frac{\big(F_x'(x, y) \big)^2}{1 - F(x, y)^2}, \\
\left(\varphi_y'(x, y) \right)^2 = \frac{\left(F_y'(x, y) \right)^2}{1 - F(x, y)^2}.
\end{aligned}
\label{eq8}
\end{equation}
Equations~(\ref{eq8}) are symmetric under the replacement $x \leftrightarrow y$, 
leading to identical process of solution  for both equations. Below, we present a solution for 
the $\varphi_x'$   only.

In expression   $\varphi_x'$, the coordinate $y$ is treated as a parameter. 
Setting it to a specific value $y = \hat{y}$ turns the functions $G$, $A$, $B$, $F$, 
and $\varphi$ into functions of the single variable $x$. In this case, the following 
notation  will be used: $\hat{f}(x) = f(x, \hat{y})$. The functions 
$\hat{F}(x)$ and $\hat{\varphi}(x)$ are defined over the interval 
$x_{\min} \leq x \leq x_{\max}$, where $x_{\min}$ and $x_{\max}$ are 
values taken from the boundary of the interferogram, and the hat symbol 
indicates that the second variable, $y$, is fixed. A version of Eq.~(\ref{eq8}) 
written for  $\varphi_x'$ as a function of the single variable $x$ reads
\begin{equation}
\big(\hat{\varphi}'(x)\big)^2 = 
\frac{\left(\hat{F}'(x)\right)^2}{1 - \hat{F}(x)^2},
\label{eq9}
\end{equation}
where, the prime ${}'$ denotes the ordinary derivative with respect to the single chosen 
variable.  To represent the phase profile from the interferogram, we construct 
$K$ different functions $\hat{F}(x)$ taken from the interferogram pattern for 
a set $\hat{Y} = \{ \hat{y}_k : 1 \leq k \leq K \}$ of values satisfying 
$y_{\min} \leq \hat{y}_k \leq y_{\max}$. In this case, Eq.~(\ref{eq9}) generates 
a series of phase profiles (slices) denoted as $\hat{\varphi}_k(x) = \varphi(x, \hat{y}_k)$, 
which represent the surface of the interferogram. Solving Eq.~(\ref{eq9}) for 
these profiles requires a corresponding boundary condition at $x = x_{\min}$, 
specifically $\varphi(x_{\min}, \hat{y}_k)$. 

Note that, in the general case, the 
boundaries of each slice may vary depending on the selected position within 
the interferogram. This leads to the more general boundary contour 
$\mathcal{D}$  and defines position-dependent limits $x_{\min}(\hat{y}_k)$ 
and $x_{\max}(\hat{y}_k)$ for each slice indexed by $\hat{y}_k$ (see Analytical Example). 

For simplicity, in this section, we align $\mathcal{D}$ with the rectangular 
canvas of the interferogram image, ensuring that the slice boundaries remain 
spatially constant across all values of $\hat{y}_k$ (note: this refers to spatial constancy, not phase constancy).

To solve Eq.~(\ref{eq9}), we first   apply the relation $\sqrt{f^2} = |f|$, where $|\cdot|$ 
denotes the modulus, and $|f(x)| = \operatorname{sgn}(f) \cdot f \geq 0$, with 
the symbol $\operatorname{sgn}(\cdot)$ standing for the sign function 
returning $\pm1$. Then, integrating Eq.~(\ref{eq9}), we obtain
\begin{equation}
\hat{\varphi}(x) = \Phi_{0\hat{y}_k} + \operatorname{sgn}\bigl(\hat{\varphi}'(x)\bigr) 
\int_{x_{\min}}^{x} \!\!\! \mathbf{K}(\xi) \, d\xi.
\label{eq10}
\end{equation}
where we introduce the shortcut notation for the integrand 
\begin{equation}
\mathbf{K}(x) = |\hat F'(x)|/\sqrt{1 - \hat{F}(x)^2}\;,
\label{eq16}
\end{equation}
and  $\Phi_{0\hat{y}_k}$ denotes the boundary value $\varphi(x_{\min},\hat{y}_k)$ 
computed from the boundary function 
$\Phi_{0y}(y)=\varphi(x_{\min},y)$ at $y=\hat{y}_k$;  the index $0$ 
indicates the minimal value, $x_{\min}$. According to 
the boundary ${\cal D}$ definition, $\Phi_{0y}(y)$ can be found from second equation in 
Eqs.~(\ref{eq8})  by solving it along the $y$-axis for $x = x_{\min}$. 
The solution for $\Phi_{0y}(y) = \varphi(x_{\min}, y)$  is similar to Eq.~(\ref{eq10}) and reads
\begin{equation}
\!\!\!\Phi_{0y}(y) =\Phi_{00} + \operatorname{sgn}
\bigl(\varphi'(x_{\min}, y)\bigr) \!\!\!  \int_{y_{\min}}^{y} 
\!\!\!\!\!\!   \mathbf{K}(x_{\min}, \psi) \, d\psi,
\label{eq11}
\end{equation}
where, $\Phi_{00}=\varphi(x_{\min}, y_{\min})$ with the same 
convention for indices marking the minimal values $x_{\min}$ and $y_{\min}$, 
respectively. Eq.~(\ref{eq11}) defines the boundary conditions for Eq.~(\ref{eq10}) 
with $y = \hat{y}_k$. The condition $\Phi_{00}$ at this point can be determined 
from Eq.~(\ref{eq4}) written in the Cartesian coordinate system
\begin{equation}
\cos \Phi_{00} = F_{00},
\label{eq12}
\end{equation}
where, $F_{00} = F(x_{\min}, y_{\min})$. From Eq.~(\ref{eq7}), $F_{00} = (G_{00} - A_{00})/B_{00}$, 
where we employ the notation  $f_{00}=f(x_{\min}, y_{\min})$. Eq.~(\ref{eq12}) has two 
solutions for $\Phi_{00}$ in the interval $-\pi \leq \Phi_{00} \leq \pi$. 
While both solutions satisfy the initial $G(x, y)$ pattern, only one of them corresponds to  the 
selected phase direction. Since $\Phi_{00}$ is a constant, it shifts the 2D phase profile $\varphi(x, y)$ as a whole. Thus, either solution of Eq.~(\ref{eq12}) will not affect the general shape of the phase.

The function $\operatorname{sgn}\bigl(\hat{\varphi}'(x)\bigr)$ defines whether the 
phase increases or decreases within each interval along $x$ bounded by  extrema of $\hat{\varphi}(x)$. 
In general, these extrema are identified as the subset of roots of $\hat{\varphi}'(x) = 0$. 
However, since  $\hat{\varphi}'(x)$ is not directly accessible, 
we instead rely on $(\hat{\varphi}'(x))^2$ from Eq.~(\ref{eq9}), 
whose both sides simultaneously vanish at the critical points of $\hat{\varphi}(x)$, leading with Eq.~(\ref{eq16}) to the condition:
\begin{equation}
\mathbf{K}^2(x)= 0,
\label{eq13}
\end{equation}
solving Eq.~(\ref{eq13}) yields the roots corresponding to critical points of $\hat{\varphi}(x)$. 
These roots must then be classified into extrema and non-extrema (e.g., flat inflection points). 
This classification relies on the behaviour of $\mathbf{K}$ in proximity of each root, but because 
$\mathbf{K}(x)=|\hat{\varphi}'(x)|$  is always positive, it inherently limits the ability of $\mathbf{K}$ to detect the nature of roots having non-obvious properties.

The inability to distinct nature of the roots by 
using only $\mathbf{K}$ function reflects the single-interferogram fundamental 
degeneracy with respect to the non-obvious root identification. To bypass this 
limitation, we restrict our attention to a subclass of phase functions whose 
critical points are all extrema $\mathcal{P}_{\mathrm{ext}}$, 
reserving $\mathcal{P}_{\mathrm{ext}}^{+}$ for the extended class 
including non-obvious roots. This constraint allows full recovery from 
a single interferogram without additional assumptions (see Discussion section).

For $n$ roots of Eq.~(\ref{eq13}) representing $n$ extrema $\chi_i$, $1 \leq i \leq n$, the whole interval should be divided into a sequence of $n+1$ segments $s_{x,i} : \{ \chi_{i-1} \leq x \leq \chi_i \}$, $1 \leq i \leq n+1$, where each 
segment is characterized by a specific value of sign  $\sigma_{x,i} = \operatorname{sgn}\bigl(\hat{\varphi}'(x)\bigr)$ 
with $x \in s_{x,i}$. For both $s_{x,i}$ and $\sigma_{x,i}$, the first index 
denotes the axis of the solution and the second one is the number of extremum. 
The sign alternates between the adjacent segments. Thus, the whole sign 
sequence is determined by the first sign  $\sigma_{x,1}$ in the first segment 
$s_{x,1} = \{  x_{\min} \leq x \leq \chi_1 \}$. Two possible values of 
$\sigma_{x,1} = \pm 1$ correspond to  solutions for two opposite 
phase shapes — a result of the quadratic form of Eq.~(\ref{eq9}), 
exhibiting another type of the single-interferogram degeneracy, related to the global phase sign (see Discussion section).

During practical computation for the selected slice $k$, it is convenient to apply  the integral of 
Eq.~(\ref{eq10}) per segment $s_{x,i}$ while the 
argument $x \in s_{x,i}$. Then, the phase at the endpoint $\chi_{i-1}$ 
of the preceding segment, $\Phi_{i-1,\hat{y}} = \varphi(\chi_{i-1}, \hat{y})$, 
must be added to the integral value. Thus, for the interval $\chi_{i-1} \leq x \leq \chi_i$, 
the phase $\hat{\varphi}_i(x)$ reads
\begin{equation}
\hat{\varphi}_i(x) = \Phi_{i-1,\hat{y}} + 
\sigma_{x,i} \int_{\chi_{i-1}}^{x} \!\!\!
\mathbf{K}(\xi) \, d\xi.
\label{eq14}
\end{equation}
The final phase $\hat{\varphi}(x)$ is a union of all segment-wise 
phases $\hat{\varphi}_i(x)$ over all segments 
\begin{equation}
\hat{\varphi}(x) = \bigcup_{i=1}^{n +1} \hat{\varphi}_i(x) 
\label{eq15}
\end{equation}

Summarize  the procedure for  the phase computation  along the $x$-axis.
The process 
begins with obtaining the interferogram function $F(x, y)$, from which $K$ slices 
$\hat{F}(x)$ are extracted. For each slice, Eq.~(\ref{eq13}) is solved, the set 
of extrema $\chi_{i}$ is obtained, and sign segments $s_{x,i}$ are defined. 
An initial sign $\sigma_{x,1}$ is then assigned arbitrarily, and the full sequence 
of signs $\sigma_{x,i}$  is derived. The same value of $\sigma_{x,1}$ is used 
consistently for all slices. Next, Eq.~(\ref{eq14}) is applied for each slice 
 to compute the corresponding phase $\hat{\varphi}(x)$. 
The initial  condition $\Phi_{00}$ remains the same for all slices, while the 
boundary condition curve $\Phi_{0y}(y)$ must be found via Eq.~(\ref{eq11}) 
and used for each slice $\hat{\varphi}(x)$ with respect to the given $\hat{y}$. 

After reconstructing all slices, the final phase profile is examined. If the overall shape 
appears mirrored relative to expectations, the sign $\sigma_{x,1}$ should  be inverted 
to correct the orientation along the $x$-axis. Alternatively, the 
entire phase shape  can be mirrored, which is often a more practical solution.

For the solution along the $y$-axis, the variable $x$ is replaced by $y$ in the above 1D 
solution, while selecting $x = \hat{x}_m,\ (1 \le m \le M)$. This yields 
a stack of $M$ phase profiles 
$\hat{\varphi}_m(y) = \varphi(\hat{x}_m, y)$. These slices require conditions 
of the boundary function $\Phi_{x0}(x)$ for $x=\hat{x}_m$, representing 
$ \varphi(\hat{x}_m, y_{\min})$ along the $x$-axis, ultimately leading 
to the same initial condition $\varphi(x_{\min}, y_{\min}) = \Phi_{00}$ 
determined from the interferogram, as described above. The $y$-axis has its 
own set of extrema $\gamma_i$, providing the segments $s_{y,i}$ and 
their corresponding signs $\sigma_{y,i}$. The sign  $\sigma_{y,1} = \pm 1$ 
must be chosen initially, and a trial with the opposite choice of $\sigma_{y,1}$ 
for the $y$-axis may be needed to match the expected orientation of the reconstructed phase.

In general, for each point $(\hat{x}_m, \hat{y}_k)$ in the interferogram, our method 
delivers two orthogonal components of the phase profiles $\hat{\varphi}_k(x)$ 
and $\hat{\varphi}_m(y)$ by solving two independent  ODEs, 
effectively representing the  2D phase surface. 

From Eq.~(\ref{eq9}), one might suspect an indeterminate form $0/0$ near the 
extrema of the interferogram function $F$, where $F \to \pm 1$ and $F' \to 0$. 
However, the quotient on the right-hand side of Eq.~(\ref{eq9}) remains 
finite and does not diverge. Indeed, Eq.~(\ref{eq4}) shows that $1 - F^2 = \sin^2 \varphi$, 
while both partial derivatives $F'_x$ and $F'_y$ contain a factor of $\sin \varphi$. 
As a result, the squared terms $(F'_x)^2$ and $(F'_y)^2$ are also proportional 
to $\sin^2 \varphi$, and the apparent singularity cancels out. The same conclusion applies to Eq.~(\ref{eq6}).

A similar concern arises in the numerical evaluation of the integrand $\mathbf{K}(x)$ 
in Eqs.~(\ref{eq10}), (\ref{eq11}), and (\ref{eq14}), where a $0/0$ form may 
occur for the same reasons discussed above. In this case, applying l’Hospital’s rule 
to $\mathbf{K}(x)$ as defined in Eq.~(\ref{eq16}) shows that terms contributing 
to divergence cancel out, leaving a finite result represented by the square root of 
the sign-specific second derivative of $\hat{F}$, where the sign of the second 
derivative depends on the extremum value of $\hat{F}(x)$, making the square 
root argument always positive and suitable for numerical evaluation even at the extrema of $F$.

\section{Examples with Unmodulated Phase}
\subsection{Analytical Example. Parabolic Phase Constrained at Boundary, $\varphi|_D=0$}

Recovering the phase via the first-order differential equation in Eq.~(\ref{eq9}) provides 
a fully analytical demonstration, impossible with traditional discrete approaches. To our knowledge, such a direct analytical phase retrieval from a single interferogram was never reported before.  

We proceed as follows: first, we define a known seed 
phase distribution $\phi$ over the spatial domain, next, we generate the 
corresponding interferogram represented by $G$, from which we 
construct the interferogram function $F$ by using Eq.~(\ref{eq3}). 
Finally, we solve Eq.~(\ref{eq9}) and compare the retrieved 
phase $\varphi$ with the original seed phase $\phi$.

\begin{figure}[t]
\centering
\includegraphics[width=0.7\linewidth]{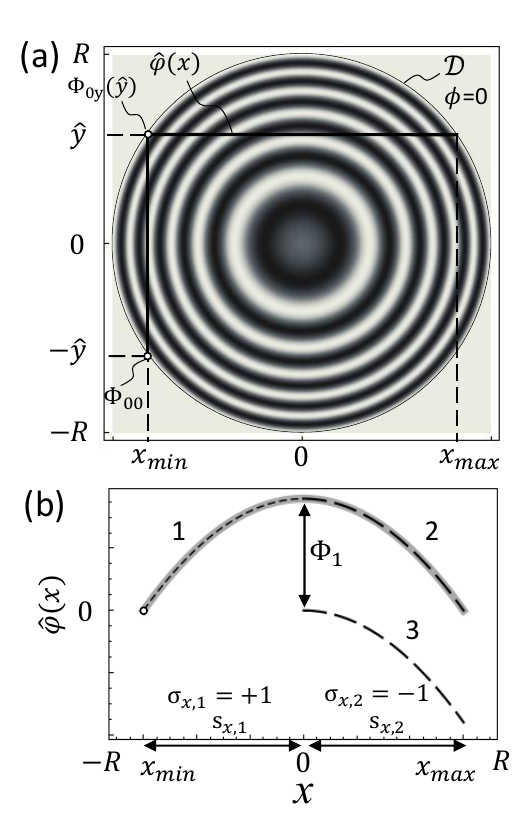}
\caption{Analytical example of phase recovery. 
Panel~(a) shows a Newton-type 
interferogram representing a parabolic fringe pattern constrained by a 
circular domain ${\cal D}$ with zero phase $\phi|_{\cal D} = 0$. The 
chord $\hat{y}$ represents the phase recovery path; the boundary 
conditions $\Phi_{0y}(\hat{y})$ and $\Phi_{00}$ are marked by white points. 
Panel~(b) illustrates the recovery process: black dashed curves~1 and~2 
represent the phase segments $\hat{\varphi}_{s_{x,1}}(x)$ and $\hat{\varphi}_{s_{x,2}}(x)$ 
corresponding to the integration segments $s_{x,1}$ and $s_{x,2}$, respectively. 
The black dashed curve~3 represents the integral in Eq.~(\ref{eq18}). The thick gray 
parabolic curve shows the resulting phase $\hat{\varphi}(x)$.}
\label{fig1}
\end{figure}

Consider the seed phase defined as an even parabolic function, $\phi(x,y) = R^2 - x^2 - y^2$ 
with the boundary $\mathcal{D} = \{ (x,y) : R^2 - x^2 - y^2 = 0 \}$, 
where $R$ is constant. The phase is constrained as $\phi = 0$ at the boundary 
curve. Consider the 1D phase recovery task for an 
arbitrary chord $-R \leq \hat{y} \leq R$ limited  by
$x_{\min} = -\hat{R}$ and $x_{\max} = \hat{R}$, 
where $\hat{R} = \sqrt{R^2 - \hat{y}^2}$, see  Fig.~\ref{fig1}(a).

For constructing an interferogram, assume a flat reference wave with 
unit amplitude  and zero phase at the detector plane, $a_r = 1$ 
and $\phi_r = 0$. Also assume the object wave having  a constant amplitude $a_s$
 over the interferogram, and its phase $\phi_s=\phi(x,y)$  is the seed phase defined above. Then, in Eq.~(\ref{eq2}), the coefficients $A$ and $B$ are some constants. The gray function reads
\begin{equation*}
\hat G(x) = A + B \cos\big(\hat{R}^2 - x^2 \big).
\end{equation*}

The interferogram has a zero-order fringe type and the typical pattern is 
presented in Fig.~\ref{fig1}(a). The Eqs.~(\ref{eq7}) and~(\ref{eq4}) read
\begin{equation*}
\hat{F} = \frac{\hat G(x) - A}{B}, \hspace*{2em} \hat{F} = \cos\big(\hat{R}^2 - x^2\big),
\end{equation*}
leading to the expression $\mathbf{K}(x)=|\hat{F}_x'|/\sqrt{1 - \hat{F}^2} = 2|x|$. Eq.~(\ref{eq13}) 
gives a condition for the roots $4x^2 = 0$. In the interval $-\hat{R} \leq x \leq \hat{R}$ 
there is a single root $\chi_1 = 0$, being the single  extremum point of the phase 
function, so it produces only two sign segments. In the first segment 
$s_{x,1} : \{-\hat{R} \leq x \leq 0\}$ the sign $\sigma_{x,1} = +1$, the integrand 
of Eq.~(\ref{eq14}) is  $\mathbf{K}(\xi)=2|\xi|$, and the phase $\hat{\varphi}_{s_{x,1}}(x)$ for this interval reads
\begin{equation}
\hat{\varphi}_{s_{x,1}}(x) = \sigma_{x,1} \int_{-\hat{R}}^{x} 2|\xi| \, d\xi = \hat{R}^2 - x^2.
\label{eq17}
\end{equation}

Note, the conditions at the boundary are trivial, the function $\Phi_{0y}(\hat{y}) = 0$, 
as well as $\Phi_{00} = 0$, because the phase vanishes at all points of the boundary, 
$\phi|_{\mathcal{D}} = 0$, significantly simplifying computation, see Fig.~\ref{fig1}(b).
 In the second segment $s_{x,2} : \{0 \leq x \leq \hat{R}\}$ we have 
 $\sigma_{x,2} = -1$, and the phase for this interval, $\hat{\varphi}_{s_{x,2}}(x)$, 
 is evaluated by following Eq.~(\ref{eq14})
\begin{equation}
\hat{\varphi}_{s_{x,2}}(x) = \Phi_1 + \sigma_{x,2} \int_0^{x} 2|\xi| \, d\xi = \Phi_1 - x^2,
\label{eq18}
\end{equation}
where $\Phi_1 = \hat{R}^2$ is the phase value at the end point $x=0$ of the 
preceding segment $s_{x,1}$ computed from Eq.~(\ref{eq17}). 
Uniting the segments we obtain the final phase 
\begin{equation*}
\hat{\varphi}(x) = \hat{\varphi}_{s_{x,1}}(x) \cup \hat{\varphi}_{s_{x,2}}(x) = \hat{R}^2 - x^2,
\end{equation*}
for the entire interval $-\hat{R} \leq x \leq \hat{R}$. As 
$\hat{R}^2 = R^2 - \hat{y}^2$ and $\hat{y}$ is arbitrarily selected, we 
conclude that the final phase $\hat{\varphi}(x) = R^2 - x^2 - \hat{y}^2$ 
coincides with the seed phase $\phi$ for $y=\hat y$. Fig.~\ref{fig1}(b) illustrates the phase recovery procedure.

Summarizing for the parabolic phase case, the method operates with a 
single extremum point, resulting in two intervals of integration. Note, the changes 
in the maximum phase values  affect only the number of fringes in the interferogram, 
but do not alter the number of extrema or integration segments. By setting the phase 
to zero at the boundary, consideration of $\varphi(x_{\min}, y)$ can be omitted, 
simplifying the solution. The method provides two possible solutions corresponding 
to $\sigma_{x,1} = +1$, shown in Fig.~\ref{fig1}(b), and $\sigma_{x,1} = -1$ (not shown), 
which is a mirror reflection of the first one with respect to the $x$-axis. The correct 
solution is selected by comparison with the known seed phase.

\subsection{Numerical Examples with Phase Unconstrained at Boundary, $\phi|_D$}

Several numerical interferogram models were selected for testing, each introducing a gradual increase in fringe pattern complexity to highlight different aspects of the method. In all cases, the seed phase $\phi$ remains unconstrained along the interferogram boundary. The models include:

\begin{enumerate}

\item A parabolic $\phi$, solved using two arbitrarily chosen orthogonal paths. This case contains a single root (an extremum) and shares the same sign of $\sigma_{x,1}$ and $\sigma_{y,1}$ along the $x$- and $y$-axes, respectively. The recovered phase is in $\mathcal{P}_{\mathrm{ext}}$ class; 

\end{enumerate}
The remaining examples are solved using 21 paths directed along the x-axis and uniformly distributed along the y-axis:

\begin{enumerate}[resume]

\item A hyperbolic paraboloid saddle $\phi$ with a single root (an extremum) along both the $x$- and $y$-axes, exhibiting opposite signs of $\sigma_{x,1}$ and $\sigma_{y,1}$ along these axes, respectively. The recovered phase is in $\mathcal{P}_{\mathrm{ext}}$ class;

\item A mixed quadratic-cubic saddle $\phi$ with a single root corresponding to an extremum along the $x$- and with a single root corresponding to a flat inflection along the $y$-axis, exhibiting same signs of $\sigma_{x,1}$ and $\sigma_{y,1}$ along these axes, respectively. The recovered phase is in an extended class $\mathcal{P}_{\mathrm{ext}}^{+}$;

\item A warped $\phi$ (case one), containing two roots along the $x$-axis and one along the $y$-axis (all extrema), with $\sigma_{x,1}$ and $\sigma_{y,1}$ sharing the same sign along both axes. It also includes a non-stationary inflection point along the $x$-axis, which does not correspond to a root of Eq.~(\ref{eq13}), and thus, does not affect the integration. The recovered phase is in $\mathcal{P}_{\mathrm{ext}}$ class;

\item A warped $\phi$ (case two), featuring three roots along the $x$-axis, one corresponding to a flat inflection and two to extrema, and one root along the $y$-axis (an extremum). The sing of $\sigma_{x,1}$ and $\sigma_{y,1}$ is the same for both axes. The recovered phase is in an extended class $\mathcal{P}_{\mathrm{ext}}^{+}$;

\item A Gaussian $\phi$, used as a baseline case to highlight the difference between an unmodulated seed phase and a linearly modulated seed phase. It has one root (an extremum) along $x$ and $y$ axes and same signs of $\sigma_{x,1}$ and $\sigma_{y,1}$ for both coordinates. The recovered phase is in $\mathcal{P}_{\mathrm{ext}}$ class.
\end{enumerate}

These  models employ the unmodulated form of $\phi$, resulting in the appearance of zero-order (Newton-type) fringes in the interferogram. Boundary conditions are incorporated according to the described method, using the 2D solution framework. 

Unless the seed phase $\phi$ is zero at the initial point $(x_{\min}, y_{\min})$, yielding $\Phi_{00} = 0.0$, the recovered phase $\varphi$ will include a constant offset relative to $\phi$. When comparing the shapes of the seed and recovered phases, this offset is disregarded if it is negligible (i.e., much smaller than the maximum phase value) and corrected for, when significant, by shifting the seed phase accordingly.

To highlight the difference in root interpretation, in addition to the examples from the $\mathcal{P}_{\mathrm{ext}}$ class, which exhibit only extrema roots, we present two examples from the extended class $\mathcal{P}_{\mathrm{ext}}^{+}$, where extrema roots are combined with a single flat-inflection root.

To simplify the interpretation of the interferograms in all examples, the following 
assumptions are made: the interferograms result from interference between 
a flat reference wave and an object wave propagating in parallel; the 
reference wave has zero phase at the interferogram plane; the entire 
phase $\phi$ is attributed to the object wave; and the amplitudes  of 
both the reference and object waves are equal  to unity. The phase in 
all  examples was restored  numerically  using Mathematica 10.4 (Wolfram Research, Inc.).

The following generic process
was applied:

\begin{enumerate}

\item Define  the seed phase $\phi(x,y)$ in a rectangular region;

\item Create a uniform intensity envelop gray function $G(x,y)$ with $A = B = 1$, leading to 
$G(x,y) = 1 + \cos(\phi)$ for all examples;

\item Create arrays $\{x_a\}$ and $\{y_a\}$ with a constant rational increment, 
each containing 401 nodes indexed by $1 \leq a \leq 401$. Construct a 
synthetic numerical interferogram by converting $G$ into the 
2D array. Then, for a subset of 21 equally 
spaced nodes fully spanning the $y$-boundary, compute a 2D array of interferogram function $\{F_{a,b}\}$ with $b=\{1, 21, 41, ..., 401\}$, according to Eq.~(\ref{eq7}),
i.e., $y_1=y_{min}$ and $y_{401}=y_{max}$.

Note: for Example 1, instead of using the above subset, select two 
individual nodes, $\hat{x}$ and $\hat{y}$, both satisfying $1 \leq a \leq 401$, 
and construct only two orthogonal 1D arrays: $\{F_{\hat{x},a}\}$ and $\{F_{a,\hat{y}}\}$, 
aligned along the $x$- and $y$-axes, respectively.

\item Interpolate each 1D sub-array of the interferogram function in $\{F_{a,b}\}$ 
using splines along all $x$-nodes for each fixed $y_b$, 
resulting in array of 1D functions $\{F(x, y_b)\}$;

\item Using interpolation along $y$-nodes for $x=x_1$ construct  the 
function $F(x_1, y)$. Then, for $F(x_1, y)$, use Eq.~(\ref{eq13}) to 
find the roots $\{\gamma_i\}$, identify extrema, compute the segments $\{s_{y,i}\}$, 
and then set their signs 
$\{\sigma_{y,i}\}$.  Employ Eqs.~(\ref{eq14}) and (\ref{eq15}) written 
for $y$ variable to reconstruct the phase $\varphi(x_1, y)$, which serves as the 
boundary condition for reconstructing the phases $\varphi(x, y_b)$ 
for given $y_b$ along $x$;

\item Reconstruct each $\varphi(x, y_b)$: for this, set the $y$-node $y_b$ 
and for the selected $F(x, y_b)$, solve Eq.~(\ref{eq13}) to find the roots $\{\chi_i\}$, identify extrema, define 
the segments $\{s_{x,i}\}$, and then set their signs $\{\sigma_{x,i}\}$, 
keeping the same $\sigma_{x,1}$ for all $y_b$. Then employ Eqs.~(\ref{eq14}) 
and (\ref{eq15}) to obtain $\varphi(x, y_b)$ taking into account the 
boundary condition $\varphi(x_1, y_b)$; then change the $y$-node and repeat.
\end{enumerate}

Indeterminate expressions do not appear during computation, as no 
exact values of $F = \pm 1$ occurrs, this is ensured by employing 
only rational coordinates for given phase recovery examples. 

\subsubsection{ Example 1: Parabolic Phase}

\begin{figure}[t]
\centering
\includegraphics[width=0.7\linewidth]{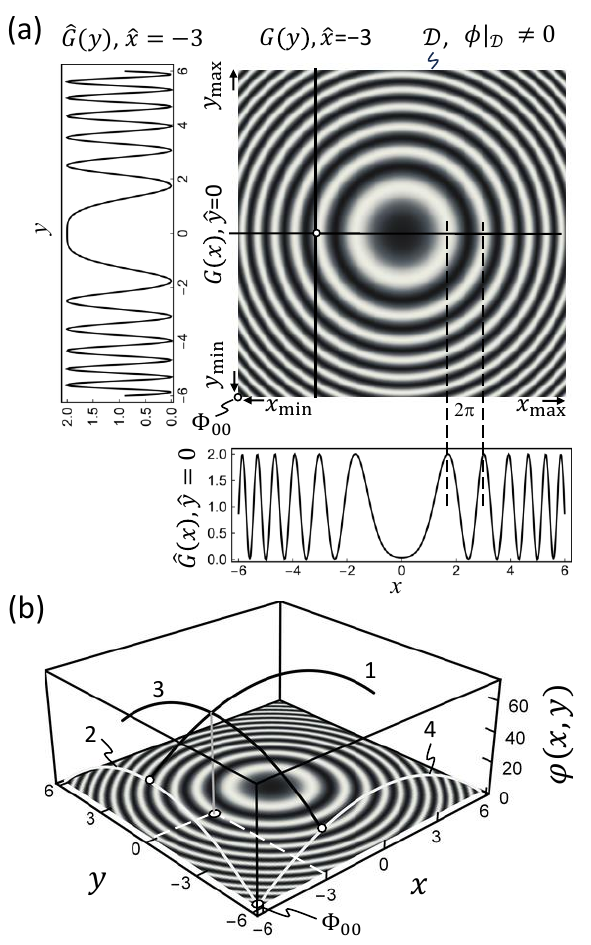}
\caption{Schematics of the phase recovery from a Newton-type interferogram representing 
a parabolic-type fringe pattern (Example 1). In (a), two functions $\hat{G}(x)$ and 
$\hat{G}(y)$ represent the cross-sections $G(x,0)$ and $G(-3,y)$, intersecting at $(-3,0)$, 
denoted by the white point in the interferogram.  ${\cal D}$ indicates the 
interferogram boundary with nonzero phase. The value $G_{00} = G(-6,-6) = 2.0$, 
used for computing $\Phi_{00}$, is shown as the white point in the lower-left corner. 
Panel (b) shows two orthogonal phase components: curve~1 for $\varphi(x,0)$ and 
curve~3 for $\varphi(-3,y)$, recovered along the $x$- and $y$-axis, respectively. 
Their intersection corresponds to the point $\bm r =(-3,0)$. The boundary conditions 
$\Phi_{0y}(y)$ and $\Phi_{x0}(x)$ are shown as white curves~2 and~4.
}
\label{fig2}
\end{figure}

Example 1 considers the parabolic seed phase $\phi(x,y) = 72 - x^2 - y^2$ with 
the square boundary ${\cal D} = \{(x,y): -6 \leq x \leq 6,\ -6 \leq y \leq 6\}$; 
the corresponding interferogram is shown in Fig.~\ref{fig2}(a). Two 1D functions 
$\hat{G}(x)$ and $\hat{G}(y)$, representing two cross-sections $G(x,0)$ and $G(-3,y)$, 
are selected for the phase recovery.

Opposite to the analytical example, where the phase boundary conditions are constrained, 
in this case the boundary conditions must be evaluated according to Eqs.~(\ref{eq11},\ref{eq12}). The first-node point $(-6,-6)$ provides the initial 
condition $\Phi_{00} = 0.0$. Two arrays having the first-node point along $x$- and $y$- 
axes define   the phase functions at the left and  bottom 
boundaries: $\Phi_{0y}(y) = \varphi(-6,y)$ and $\Phi_{x0}(x) = \varphi(x,-6)$, respectively.

Solving Eq.~(\ref{eq13}) along the $x$-axis yields the single root $\chi_1 = 0.0$ (an extremum), 
producing  two segments $s_{x,1} : \{-6 \leq x \leq 0\}$ and $s_{x,2} : \{0 \leq x \leq 6\}$. The signs of the segments, $\sigma_{x,1} = +1$ and $\sigma_{x,2} = -1$,
are selected to match the seed phase. Solving Eq.~(\ref{eq13}) along the $y$-axis yields 
also the single root $\gamma_1 = 0.0$ (an extremum), giving similar segment boundaries, and sign assignments as along the $x$-axis.

Fig.~\ref{fig2}(b) demonstrates phase recovery along $x$- and $y$- directions 
corresponding to $\varphi(-3,y)$ and $\varphi(0,x)$ phase components for the node $(-3,0)$.

\subsubsection{ Example 2: Hyperbolic Paraboloid Phase: Saddle}

Example 2 illustrates the numerical recovery of the saddle-type seed phase with 
$\phi(x,y) = x^2 - y^2$ within the square boundary ${\cal D} = \{(x,y) : -6 \leq x \leq 6,\ -6 \leq y \leq 6\}$. The recovered profiles $\varphi(x,y_b)$ produce an effective approximation of the 2D phase surface. The corresponding interferogram is shown in Fig.~\ref{fig:fig3}(a). 

The initial condition for $\varphi(x_{\min}, y_{\min})$ corresponds to $\Phi_{00} = 0.0$ at the point $(-6,-6)$, and the boundary condition for 
each of the 21 curves along the $x$-axis is given by the function $\Phi_{0y}(y)$ taken at $y=y_b$, corresponding to $\varphi(-6, y_b)$.

Eq.~(\ref{eq13}) yields a single root $\chi_1 = \gamma_1 = 0.0$ (an extremum) along both 
the $x$- and $y$- axes, resulting in two similar integration segments for each direction: $s_{x,1} = s_{y,1} = \{-6 \leq x,y \leq 0\}$ and $s_{x,2} = s_{y,2} = \{0 \leq x,y \leq 6\}$. 
These segments are assigned different signs $\sigma$ to match the reconstructed phase $\varphi$ with the original seed phase $\phi$. Specifically, 
along the $x$-axis, the segment $s_{x,1}$ has $\sigma_{x,1} = -1$, while along the $y$-axis, the segment $s_{y,1}$ has $\sigma_{y,1} = +1$.

\subsubsection{ Example 3: Mixed Quadratic–Cubic Phase: Saddle with an inflection}

\begin{figure}[t]
\centering
\includegraphics[width=0.7 \linewidth]{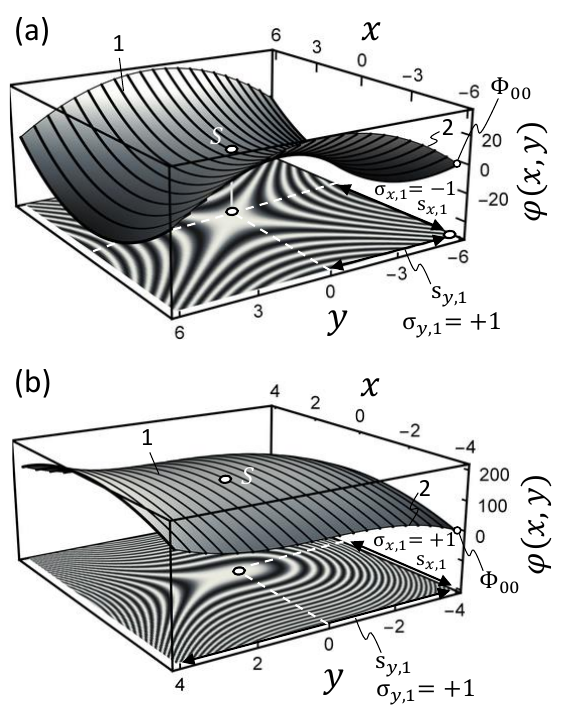}
\caption{Numerical phase recovery from a Newton-type interferogram 
representing a saddle-type fringe patterns. 
Panel~(a) depicts a saddle-type surface with the saddle point $S$ 
exhibiting identical extrema, $\chi_{1} = \gamma_{1} = 0.0$, along 
both the $x$- and $y$-axes (Example 2), whereas panel~(b) shows $S$ with an 
extremum $\chi_{1} = 0.0$ along the $x$-axis and a flat inflection $\gamma_{1} = 0.0$ along the $y$-axis  (Example 3).
For both panels (a) and (b): The gray surfaces represent the 
seed phases, with the corresponding interferograms shown at the 
bottom. The black parallel curves (denoted by 1) correspond to  
recovered phase $\varphi(x,y_b)$ overlaid with the seed phase along the $x$-axis; 
the sign $\sigma_1$ of the first segment $s_1$ is shown for both $x$- 
and $y$-axes. The boundary condition $\Phi_{0y}(y_b)$ for 
each $\varphi(x,y_b)$ are represented by curve~2; the initial condition $\Phi_{00} = 0$. 
}
\label{fig:fig3}
\end{figure}

The numerical recovery of the more complicated saddle phase profile possessing a flat inflection point along $y$-axis is illustrated by Example 3. The seed phase exhibiting a mix of the quadratic and cubic terms
$\phi(x,y) = y^3 - 5 x^2$ within the square boundary ${\cal D} = \{(x,y) : -4 \leq x \leq 4,\ -4 \leq y \leq 4\}$ and corresponding interferogram are shown in Fig.~\ref{fig:fig3}(b). 

The initial condition $\Phi_{00} = 0.513$ at the point $(-4,-4)$, and the boundary condition for 
all curves recovered along the $x$-axis is given by the function $\Phi_{0y}(y)$ taken at $y=y_b$, corresponding to $\varphi(-4, y_b)$.

Eq.~(\ref{eq13}) yields a single root $\chi_1 = 0.0$ (an extremum) along the $x$-axis, resulting in two integration segments: $s_{x,1} =  \{-4 \leq x \leq 0\}$ and $s_{x,2} = \{0 \leq x \leq 4\}$. Along $y$-axis Eq.~(\ref{eq13}) yields also a single root $\gamma_1 = 0.0$, however this is a flat inflection, and therefore must be removed, resulting to a single integration segment:  $s_{y,1} = \{-4 \leq y \leq 4\}$. The signs of the first segments are assigned to $\sigma_{x,1} = +1$ and $\sigma_{y,1} = +1$ to match shapes of the reconstructed phase $\varphi$ and the original seed phase $\phi$, Fig.~\ref{fig:fig3}(b).

In this example, the reconstructed and seed phases have a constant difference $146.266$, in Fig.~\ref{fig:fig3}(b) the seed phase is shifted accordingly to match both shapes.

\subsubsection{ Example 4: Warped Phase. Case with a non-stationary inflection}

\begin{figure}[t]
\centering
\includegraphics[width=0.6\linewidth]{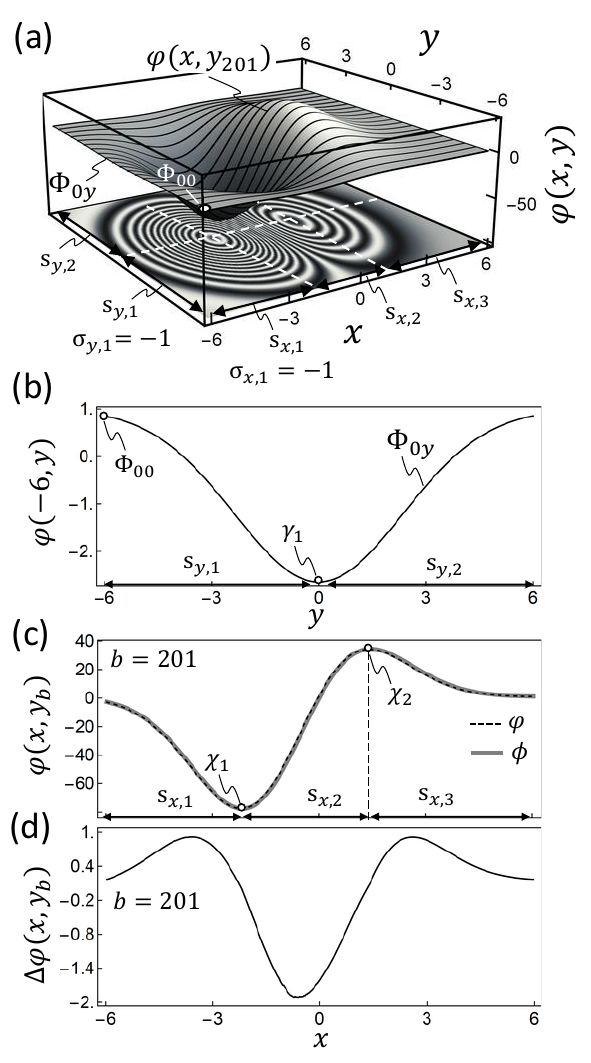}
\caption{Numerical phase recovery from a Newton-type interferogram representing a warped type of fringe pattern (case one, Example 4). 
In (a), the gray surface shows the seed phase, with the corresponding interferogram displayed beneath it. The black curves correspond to 21 recovered phase slices $\varphi(x,y_b)$ along the $x$-axis, aligning with the seed phase; the sign $\sigma_1$ of the first segment $s_1$ is indicated for both $x$- and $y$-axes. Panel (b) shows the boundary curve $\Phi_{0y}$ and the initial condition $\Phi_{00} = 0.857$, the curve has an extremum $\gamma_1 = 0.0$ providing two segments $\{s_{y,1}, s_{y,2}\}$. Panels (c) and (d) correspond to the $y$-node slice at $b = 201$: (c) shows the alignment of the seed phase $\phi(x,y_b)$ with the recovered phase $\varphi(x,y_b)$; the recovered phase has two extrema $\chi_1 = -2.182$ and $\chi_2 = 1.432$ defining three integration segments; 
(d) presents the absolute error $\Delta\varphi = \phi - \varphi$.}
\label{fig:fig4}
\end{figure}

Example 4 numerically recovers the phase from the warped type of the interferogram, Fig.~\ref{fig:fig4}(a), that corresponds to the 
seed phase $\phi(x,y) = 1 + 50 x e^{-(0.4x+0.3)^2 - (0.3y)^2}$ with 
the square boundary ${\cal D} = \{ (x, y) : -6 \leq x \leq 6,\ -6 \leq y \leq 6 \}$. 
The phase was recovered along the $x$-axis as $\varphi(x, y_b)$. The recovered phase profile, $\varphi(x,y)$, is shown in Fig.~\ref{fig:fig4}(a) as 21 curves aligned with the 2D seed phase profile $\phi$, rendered as a gray surface.

The initial condition is $\Phi_{00} = 0.857$ at the point $(-6, -6)$. The boundary conditions for each of the curves $\varphi(x, y_b)$, along the $x$-axis, 
are given by the function $\Phi_{0y}(y)$ taken at $y_b$, corresponding to $\varphi(-6, y_b)$; see Fig.~\ref{fig:fig4}(b).

There is one root $\gamma_1 = 0.0$ (an extremum) along the $y$-axis, providing two segments $s_{y,1}$ and $s_{y,2}$; the sign for $s_{y,1}$ is selected 
as $\sigma_{y,1} = -1$, Fig.~\ref{fig:fig4}(a).

There are two roots $\chi_1 = -2.182$ and $\chi_2 = 1.432$, same for all 21 curves along the $x$-axis; see the example curve $\varphi(x, y_{201})$ in Fig.~\ref{fig:fig4}(c). These two roots correspond to extrema, providing three segments $s_{x,1}$, $s_{x,2}$, and $s_{x,3}$ which are
used for integration in Eq.~(\ref{eq14}). The sign for $s_{x,1}$  is selected 
as $\sigma_{x,1} = -1$, Fig.~\ref{fig:fig4}(a). The segment $s_{x,2}$ along the $x$-axis contains a non-stationary inflection point, which is not a root of Eq.~(\ref{eq13}).

The signs $\sigma_{x,1}$ and $\sigma_{y,1}$, 
along the $x$- and $y$-axes, respectively, are selected by trial to match the seed phase $\phi$. The absolute error between the seed phase and 
the recovered phase $\Delta \varphi(x,y_b) = \phi(x,y_b) - \varphi(x,y_b)$, 
for the $y$-node 201, is presented in Fig.~\ref{fig:fig4}(d).

\subsubsection{ Example 5: Warped Phase. Case with a flat inflection}

Example 5 is a more generalized version of Example 4, recovering the phase from a warped interferogram, where each $x$-slice contains a flat inflection point. The seed phase is defined as $\phi(x,y) = 1 + 4 x^3 e^{-(0.3x+0.16)^2 - (0.3y)^2}$ within the square domain ${\cal D} = \{ (x, y) : -8 \leq x \leq 8,\ -8 \leq y \leq 8 \}$. 

\begin{figure}[t]
\centering
\includegraphics[width=0.6\linewidth]{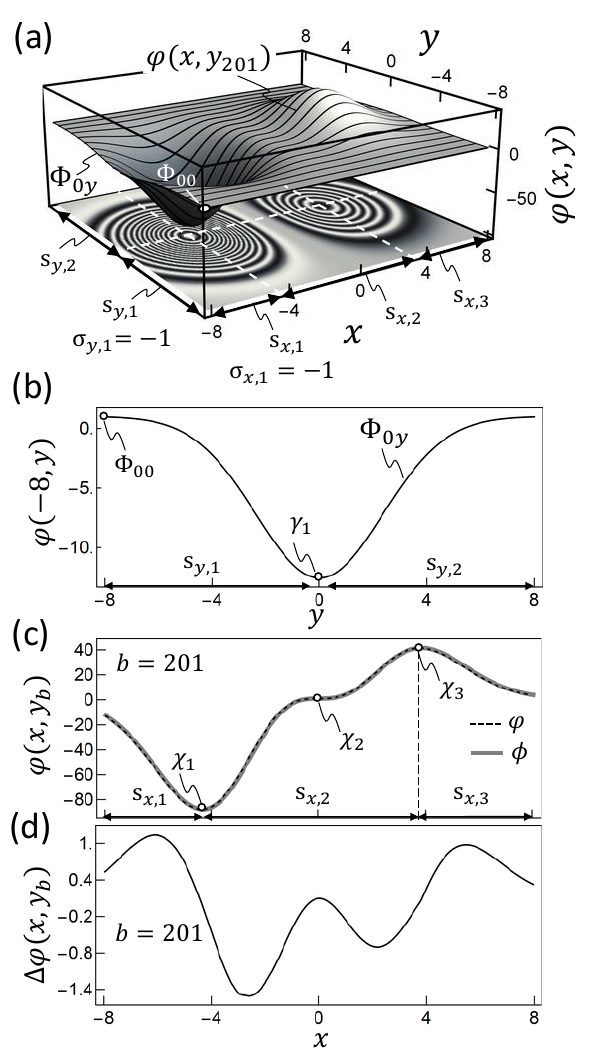}
\caption{Numerical phase recovery of the warped seed phase (case two), Example 5. 
In (a), the gray surface is the seed phase, with the corresponding interferogram beneath it. The black 
curves represent recovered phase slices $\varphi(x,y_b)$ aligned with the seed phase; the sign $\sigma_1$ of the first segment $s_1$ is shown for both $x$- and $y$-axes. In (b), the boundary curve $\Phi_{0y}$ having an extremum $\gamma_1 = 0.0$ providing two integration segments, and the boundary condition $\Phi_{00} = 0.957$ is displayed. Panels (c) and (d) correspond to the $y$-node slice at $b = 201$: (c) shows the alignment of the seed phase $\phi(x,y_b)$ with the recovered phase $\varphi(x,y_b)$; the recovered phase exhibits two extrema $\chi_1 = -4.358$ and $\chi_3 = 3.825$, and a flat inflection point  $\chi_2 = 0.0$, 
defining three integration segments; (d) represents the absolute error $\Delta\varphi = \phi - \varphi$.}
\label{fig:fig4flat}
\end{figure}

The phase recovery process is similar to Example 4. The differences include the initial condition $\Phi_{00} = 0.957$ and the boundary condition for each $x$-slice $\varphi(x, y_b)$ taken from a new boundary function $\Phi_{0y}(y)$ and evaluated at $y_b$ corresponding to the value $\varphi(-8, y_b)$, see Fig.~\ref{fig:fig4flat}(b).

As in Example 4, there is a single root $\gamma_1 = 0.0$ along the $y$-axis which corresponds to an extremum and partitions the $y$-domain into two integration segments, $s_{y,1}$ and $s_{y,2}$. The sign is assigned as $\sigma_{y,1} = -1$, following the same convention as before, see Fig.~\ref{fig:fig4flat}(a).

In contrast to Example 4, this case exhibits three roots along the $x$-axis, identical for all $x$-slices: $\chi_1 = -4.358$, $\chi_2 = 0.0$, and $\chi_3 = 3.825$. An example slice $\varphi(x, y_{201})$ is shown in Fig.~\ref{fig:fig4flat}(c). Local analysis of $\mathbf{K}(x)$ in the vicinity of each root distinguishes the extrema $\chi_1$ and $\chi_3$ from the flat inflection point $\chi_2$.

The two extrema define three integration segments, $s_{x,1}$, $s_{x,2}$, and $s_{x,3}$, which are used for integration similar to Example 4. The sign for $s_{x,1}$ is chosen as $\sigma_{x,1} = -1$, again following the convention established previously in Example 4, (Fig.~\ref{fig:fig4flat}(a)).

The absolute error between the seed phase and 
the recovered phase $\Delta \varphi(x,y_b) = \phi(x,y_b) - \varphi(x,y_b)$, for the same $y$-node 201 as in Example 4, is presented in Fig.~\ref{fig:fig4flat}(d).

\subsubsection{ Example 6: Gaussian Phase}

\begin{figure}[t]
\centering
\includegraphics[width=0.6\linewidth]{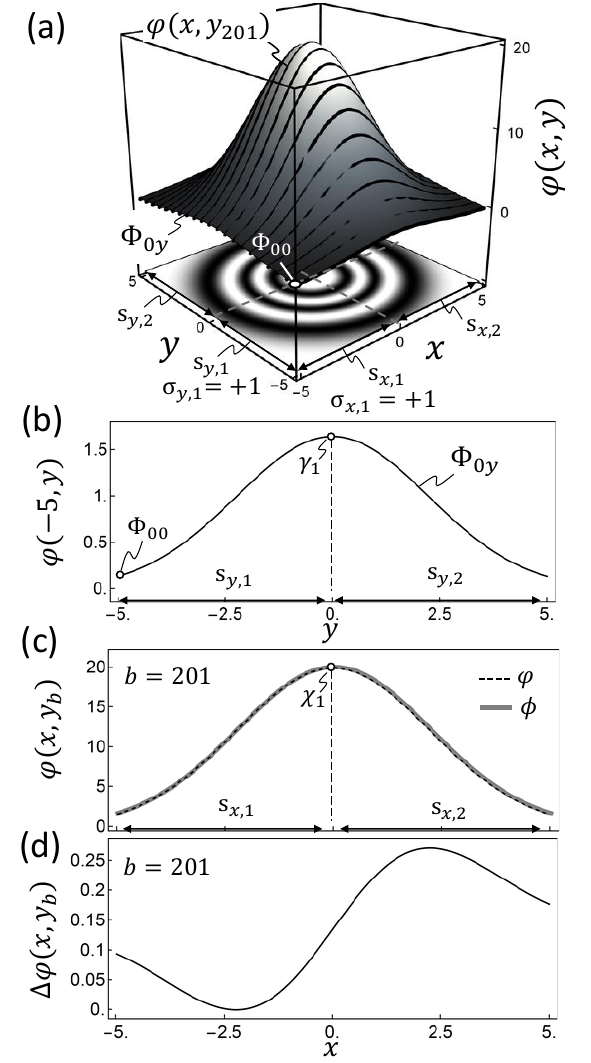}
\caption{Numerical phase recovery from a Newton-type interferogram representing a 
Gaussian-type fringe pattern (Example 6). In (a), the gray surface represents the seed phase, with the corresponding interferogram shown at the bottom. The black curves correspond to 21 recovered phase slices $\varphi(x,y_b)$ aligned with the seed phase along the $x$-axis; the sign $\sigma_1$ of the first segment $s_1$ is shown for both $x$- and $y$-axes. Panel (b) shows the boundary condition curve $\Phi_{0y}(y)$ having an extremum $\gamma_1 = 0.0$, which defines two segments $\{s_{y,1}, s_{y,2}\}$, and the  boundary condition $\Phi_{00} = 0.135$. Panels (c) and (d) correspond to the $y$-node slice at $b = 201$: (c) shows the alignment of the seed phase $\phi(x,y_b)$ with the recovered phase $\varphi(x,y_b)$; (d) presents the absolute error $\Delta\varphi = \phi - \varphi$.
}
\label{fig:fig5}
\end{figure}

Example 6 numerically recovers the phase from the Gaussian type of the interferogram, 
Fig.~\ref{fig:fig5}. The seed phase $\phi(x,y) = 20e^{-0.1(x^2 + y^2)}$ produces the 
interferogram limited by the square boundary ${\cal D} = \{(x,y) : -5 \leq x \leq 5, -5 \leq y \leq 5\}$. 
The recovered phase $\varphi(x,y_b)$ was computed along the $x$-axis, representing 
a grid of parallel curves aligned with the 2D profile of the seed phase $\phi$ rendered by the gray surface in Fig.~\ref{fig:fig5}(a).

The initial condition is $\Phi_{00} = 0.135$ at $(-5, -5)$. The boundary conditions for each of the 21 curves $\varphi(x,y_b)$ along the $x$-axis are given by the function $\Phi_{0y}(y)$ taken at  $y_b$, 
corresponding to the slices $\varphi(-5, y_b)$, as shown in Fig.~\ref{fig:fig5}(b).

Along the $y$-axis, there is one root $\gamma_1 = 0.0$ (an extremum), which defines two segments, $s_{y,1}$ and $s_{y,2}$, used for integration in Eq.~(\ref{eq14}). The sign for $s_{y,1}$ is chosen as $\sigma_{y,1} = +1$, Fig.~\ref{fig:fig5}(a). 

Similarly, along the $x$-axis, there is one root $\chi_1 = 0.0$ (an extremum), which is common to all 21 curves; see the example curve $\varphi(x, y_{201})$ in Fig.~\ref{fig:fig5}(c). This root defines two segments, $s_{x,1}$ and $s_{x,2}$, used for integration. The sign for $s_{x,1}$ is selected as $\sigma_{x,1} = +1$, Fig.~\ref{fig:fig5}(a).

Both signs $\sigma_1$ along the $x$- and $y$-axes are selected by trial to match the seed phase $\phi$. The absolute error between the seed phase and the recovered phase, $\Delta\varphi = \phi - \varphi$, for the $y$-node $b=201$ (i.e., $\varphi(x, y_{201})$), is presented in Fig.~\ref{fig:fig5}(d).

\section{ Spatially Modulated Phase }
\subsection{ Formulation and Solution}

Consider the phase that is spatially modulated along a single $x$-coordinate. 
In this case, the phase in Eq.~(\ref{eq8}) is replaced 
by $\hat{\varphi}_x \to \alpha \hat{\varphi}_x + f(x)$, where 
the function $f(x)$ modulates the initial phase $\hat{\varphi}_x$ along 
the $x$-axis, while the phase component $\hat{\varphi}_y$ along the $y$-axis 
remains unmodulated. 

An important practical case corresponds to linear modulation, where 
$f(x) = \beta_0 + \beta x$, with $\beta_0$ and $\beta$ representing a constant 
phase offset and a linear phase gradient along the $x$-axis, respectively. In this 
case, the function $G(x,y)$, given by Eq.~(\ref{eq2}), is encoded by the 
modulated phase component, producing Fizeau-type fringes in the interferogram. 
In practice, such phase modulation can be achieved by making the reference and 
object beams non-parallel, either by tilting one of the mirrors forming the beams 
or by inserting a wedge of transparent material into the object beam.

The phase-retrieving equation Eq.~(\ref{eq9}) corresponding to this spatially modulated phase becomes
\begin{equation}
(\alpha \varphi'(x) + \beta )^2 =  
\frac{\left(\hat{F}'(x)\right)^2}{1 - \hat{F}(x)^2}, 
\label{eq19}
\end{equation}
Previously,  solution of Eq.~(\ref{eq9}) 
required the search and analysis of the roots $\chi_i$ of 
Eq.~(\ref{eq13}), corresponding to extrema of $\hat{\varphi}(x)$ and leading 
to the function $\operatorname{sgn}\bigl(\hat{\varphi}'(x)\bigr)$, which defines a 
particular solution of $\hat{\varphi}(x)$ on the interval $[x_\mathrm{min}, x_\mathrm{max}]$. 
In the current case, solving Eq.~(\ref{eq19}) we demand no roots $\chi_i$ exist in the interval.
Therefore, we seek the conditions under which the 
function $\operatorname{sgn}\bigl(\alpha \hat{\varphi}'(x) + \beta \bigr)$ remains constant over the entire interval $[x_\mathrm{min}, x_\mathrm{max}]$. 
This is possible by appropriately balancing the gradient of the phase $\hat{\varphi}'(x)$ with the parameters $\alpha$ and $\beta$.  

Two possible values of the sign function, $-1$ or $+1$, are denoted here as a single sign $\sigma_{x,1} = \pm 1$, corresponding to a single segment 
of integration $s_{x,1}$ for $x \in [x_\mathrm{min}, x_\mathrm{max}]$. This corresponds to the inequality
\begin{equation}
\alpha \hat{\varphi}'(x) + \beta > 0 \quad \text{or} \quad \alpha \hat{\varphi}'(x) + \beta < 0,
\label{eq20}
\end{equation}
which must hold over the entire interval $[x_\mathrm{min}, x_\mathrm{max}]$. 
Both values of $\sigma_{x,1} = \pm 1$ must be tested to determine which 
one corresponds to the actual seed phase.  

Under these conditions, Eq.~(\ref{eq10}) simplifies to a single integral along 
the $x$-axis for the selected slice at $y = \hat{y}_k$:
\begin{equation}
\hat \varphi(x) = \Phi_{0\hat{y}_k} + \sigma_{x,1} 
\int_{x_\mathrm{min}}^x 
\mathbf{K}(\xi)\,d\xi.
\label{eq21}
\end{equation}
Here, $\Phi_{0\hat{y}_k}$ is the boundary value $\varphi(x_\mathrm{min}, \hat{y}_k)$. 
This value is found from Eq.~(\ref{eq11}), requiring computation of a number of 
integrals $\varphi_i$  related to $m$ extrema $\gamma_i$ obtained by solving 
Eq.~(\ref{eq13}) along the $y$-axis
\begin{equation}
\varphi_i(x_\mathrm{min}, y) = \Phi_{00} + \sigma_{y,i} \int_{\gamma_{i-1}}^{y} 
\mathbf{K}(x_\mathrm{min}, \psi) \, d\psi.
\label{eq22}
\end{equation}
Then, the complete phase profile along the $y$-axis boundary slice at $x_\mathrm{min}$ reads
\begin{equation}
\varphi(x_\mathrm{min}, y) = \bigcup_{i=1}^{m +1} \varphi_i(x_\mathrm{min}, y),
\label{eq23}
\end{equation}
defining the boundary function $\Phi_{0y}(y)$, and consequently, for 
the selected $y = \hat{y}_k$ giving the value $\Phi_{0\hat{y}_k}$ in Eq.~(\ref{eq21}).

\subsection{ Recovery of Spatially Modulated Phase }

\begin{figure}[t]
\centering
\includegraphics[width=0.6\linewidth]{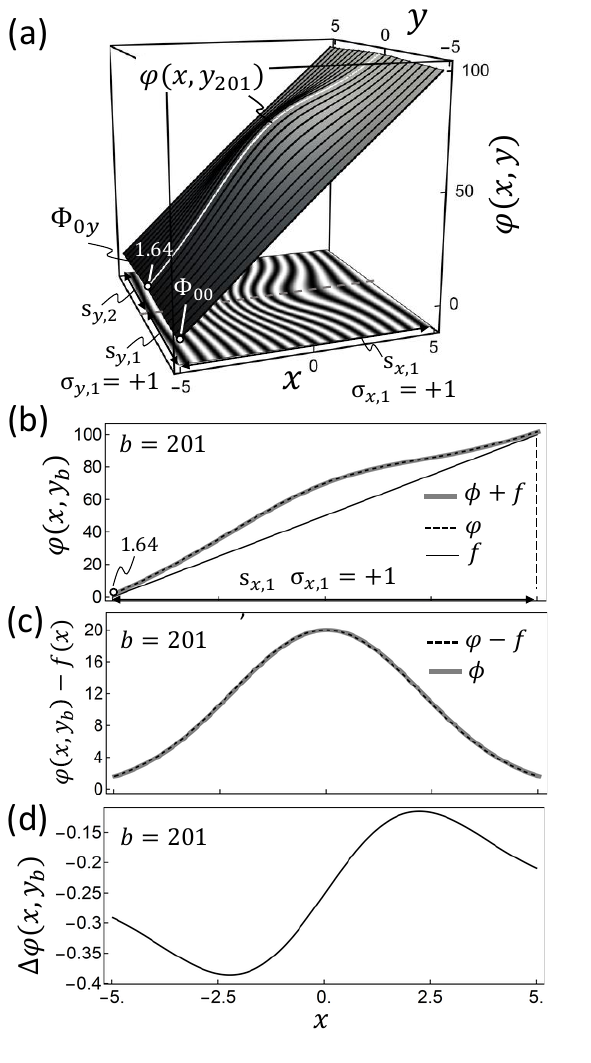}
\caption{Numerical phase recovery from a Fizeau-type interferogram representing 
linear modulation along the $x$-axis of a Gaussian-type phase (Example 7). 
In (a), the inclined gray surface shows the total seed phase, with the 
corresponding interferogram at the bottom. The black curves denote the recovered phase 
slices $\varphi(x,y_b)$ aligned with the total seed phase; the sign $\sigma_1$ 
of the first segment $s_1$ is shown for both $x$- and $y$-axes. The panels (b)--(d) 
correspond to the $y$-node slice at $b=201$: (b) shows the recovered phase 
$\varphi(x,y_b)$ together with the seed phase components: $\phi(x,y_b)$ and $f(x)$; 
(c) shows the alignment of the unmodulated seed term $\phi(x,y_b)$ with the difference $\varphi(x,y_b)-f(x)$; 
and (d) presents the absolute error $\Delta\varphi = (\phi + f) - \varphi$.}
\label{fig:fig6}
\end{figure}

Example 7 demonstrates the numerical recovery of the phase from a 
Gaussian-type interferogram with a linear spatial modulation applied 
along the $x$-axis, Fig.~\ref{fig:fig6}. The total seed phase is composed of two terms: 
the first one, representing the seed of the object beam, is the same as in Example 6 
and is given by $\phi(x,y)=20e^{-0.1(x^2+y^2)}$; the second one introduces 
the one-dimensional modulation phase $f(x)=10(x+5)$. The total seed phase 
produces a Fizeau-type fringe pattern, shown in Fig.~\ref{fig:fig6}.

Because the method, like the interferogram itself, is insensitive to an overall 
constant phase shift, the total phase at the left boundary $x=-5$ in Fig.~\ref{fig:fig6} is 
constrained to lie within a single fringe, i.e., the interval $[0,2\pi]$, to allow consistent 
comparison with Example 6. This constraint corresponds to the term $(5+x)$ 
in the modulation phase. The coefficient 10 ensures a sufficiently steep slope of the 
total phase, preventing the appearance of roots in Eq.~(\ref{eq13}) along the $x$-axis, 
as required by the inequalities in Eq.~(\ref{eq20}).

The interferogram is defined within the same square domain used in Example 6. 
The phase $\varphi(x,y_b)$ is recovered along the $x$-axis for 21 equally spaced slices, forming a grid of parallel curves that follow the 2D profile of the 
total seed phase $\phi(x,y)+f(x)$, as shown by the gray surface in Fig.~~\ref{fig:fig6}(a). 
The boundary condition $\Phi_{00}$ is $1.64$, while the boundary 
curve $\Phi_{0y}(y)$ is computed for $x=-5$. Along the $y$-axis, there 
is a single root at $y_1=0.0$ (an extremum), which defines two integration segments 
$\{s_{y,1},s_{y,2}\}$ according to Eq.~(\ref{eq13}). For the first segment 
$s_{y,1}$, the sign $\sigma_{y,1}=+1$ is used, similar to Example 6, see Fig.~\ref{fig:fig5}(b).

In contrast to Example 6, no roots are present along the $x$-axis, leading to a 
single integration segment $s_{x,1}$ with the sign $\sigma_{x,1}=+1$, see Fig.~\ref{fig:fig6}(a, b). 
The recovered phase profile corresponding to the central node $b=201$, shown as the white 
curve in Fig.~\ref{fig:fig6}(a), demonstrates good agreement between the total seed phase 
$\phi(x,y_{201})+f(x)$ and the recovered phase $\varphi(x,y_{201})$, as illustrated in Fig.~\ref{fig:fig6}(b).

Subtracting the modulation function from the recovered phase, i.e., computing $\varphi(x,y)-f(x)$, 
yields the phase that closely matches the original (unmodulated) object beam seed 
phase $\phi(x,y)$, shown in Fig.~\ref{fig:fig6}(c). The absolute error between 
the total seed phase and the recovered phase, defined as $\Delta\varphi=(\phi+f)-\varphi$, 
is plotted for the $y$-node at $b=201$ in Fig.~\ref{fig:fig6}(d). In both $x$- and $y$-axes, 
the signs $\sigma_1$ are selected by trials to match the original seed phase $\phi$ of the object beam.

\section{Discussion }

From a methodological perspective, the proposed continuous process of phase-retrieving  consists of three 
main parts: (i) obtaining oscillating function $F$, with a uniform intensity envelope bounded by the interval $[-1,1]$ serving as the input data; (ii) computing non-oscillating function $\mathbf{K}$, solving Eq.~(\ref{eq13}), and identifying extrema, this step determines the sequence of integration segments where the phase derivative changes sign; (iii) recovering the phase by direct integration of $\mathbf{K}$ along  identified segments taking into account boundary conditions.

The continuous formulation allows either analytical or numerical implementation, depending on interferogram complexity.

It is noteworthy that the identification of  extrema, i.e., the segments $s_i$ 
associated with specific signs of the phase derivative $\sigma_i$ is performed 
prior to the actual phase reconstruction. Since this step requires only the function 
$F$, it can be treated as an independent tool that provides both the locations 
and the signs of phase derivative directly from the interferogram. This 
feature can be particularly valuable for heuristic, piecewise phase unwrapping 
algorithms, which rely on such segmentation.

Once the sequence of extrema is determined, the subsequent integration along each segment $s_i$, with proper accounting for the corresponding 
sign $\sigma_i$, is straightforward, provided that any coordinates resulting in $0/0$ indeterminacies in the integrand of Eq.~(\ref{eq16}) are isolated and treated separately.

Several factors may further complicate the implementation of the proposed method. 
These include: the inherent degeneracies of the reconstruction problem when 
using a single interferogram; the necessity of selecting an appropriate interference model;  identification of the roots; 
and non-uniformities in the interferogram intensity arising 
from uneven illumination or noise. These complications are addressed below.

\subsection{ Global Phase Sign Degeneracy}

It is well known that a single interferogram exhibits phase reconstruction degeneracy 
with respect to the direction—either toward or away from the interferogram plane. 
This degeneracy originates from Eq.~(\ref{eq4}), where the cosine function is even in $\varphi$, i.e., $\cos\varphi = \cos(-\varphi)$. As a result, two opposite phase profiles produce the same interferogram.

In our method, a similar ambiguity arises from the quadratic form of the phase-retrieving differential equations Eqs.(\ref{eq6}), leading to an uncertainty in the sign $\sigma_1$, the first term in the alternating sequence of signs associated with the integration intervals $s_i$. The two possible 
values, $\sigma_1 = \pm 1$, correspond to phase profiles that are mirror reflections of each other with respect to the interferogram plane.

When reconstructing a 2D phase profile, as shown in the numerical examples, both coordinate directions are involved: the $y$-axis is used to define boundary conditions, while the $x$-axis is used for actual 
phase reconstruction. In this case, the ambiguity appears independently in the first-segment signs $\sigma_{x,1} = \pm 1$ and $\sigma_{y,1} = \pm 1$. 
However, the total number of distinct phase solutions remains two, since the first segments $s_{x,1}$ and $s_{y,1}$ must correspond to the same 
underlying 2D phase profile. Only one consistent choice of the sign pair $(\sigma_{x,1}, \sigma_{y,1})$ is valid for each solution.

Interestingly, this global sign degeneracy makes the individual identification of extrema types in Eq.~(\ref{eq13}) unnecessary for our method, thereby simplifying its implementation. The method only requires the natural alternation between maxima and minima, as any incorrect assignment is ultimately compensated by a global sign inversion in the reconstructed phase.

This type of phase reconstruction degeneracy represents an inherent limitation for any analysis of a single interferogram. However, in many practical 
applications, such as in fluid dynamics or colloidal science, the overall phase direction or the phase value at a reference point is known from the experimental context. In such cases, our method can accurately recover the continuous phase profile, 
with the ambiguity resolved by the available experimental information.

\subsection{ Identifying Roots of $\mathbf{K}$}

The solution of the phase-retrieval equation, Eq.~(\ref{eq8}), requires knowledge of $\operatorname{sgn}\bigl(\hat{\varphi}'(x)\bigr)$, as indicated by the integral in Eq.~(\ref{eq10}). A standard approach is to determine all points $x$ where $\hat{\varphi}'(x) = 0$ and thereby partition the integration domain of Eq.~(\ref{eq10}) into subintervals $\{s_{x,i}\}$ within which the sign of $\hat{\varphi}'(x)$ remains constant. For smooth functions $\hat{\varphi}$, these signs alternate strictly between consecutive subintervals, enabling one to avoid explicit computation of $\hat{\varphi}'(x)$ on arbitrary segments while controlling the initial sign $s_{x,1}$, which can take only the values $\pm 1$. Once the sequence $\{s_{x,i}\}$ is identified, by any suitable method, Eq.~(\ref{eq10}) can be integrated to recover the phase, up to a sign of the global shape.

Since $\hat{\varphi}'(x)$ cannot be evaluated directly, we propose to solve $|\hat{\varphi}'(x)| = 0$, which reduces to $\mathbf{K}(x) = 0$ and subsequently to Eq.~(\ref{eq13}). Extrema among the roots $\chi_i$ of Eq.~(\ref{eq13}) partition the integration domain into subintervals $\{s_{x,i}\}$ within which the sign of $\hat{\varphi}'(x)$ remains constant. The general challenge is to distinguish these extrema roots from the others. The remaining roots correspond to \textit{flat inflections}, where $\hat{\varphi}'(\chi) = \hat{\varphi}''(\chi) = 0$ and the first nonzero derivative is of odd order $n \ge 3$, i.e., $\hat{\varphi}^{(n)}(\chi) \ne 0$. At a flat inflection, $\hat{\varphi}'(x)$ does not change sign. Consequently, if such a root is mistakenly classified as an extremum, the sign sequence in Eq.~(\ref{eq10}) will flip from that point onward, leading from this point to an incorrect phase profile.

At present, we do not have a general method capable of separating extrema from flat inflections for arbitrary phase profiles. Instead, we briefly present below a set of observations that may assist in performing such a separation in certain cases of potential experimental relevance.

First, if there is conclusive evidence that all roots correspond to extrema, the problem described above is eliminated, and the phase can be assigned to the $\mathcal{P}_{\mathrm{ext}}$ class. This condition should be used whenever applicable. Such evidence can be found, for example, through symmetry analysis of the interferogram, particularly when the total number of roots is small. This approach appears to work more reliably for Fizeau-type interferograms than for Newton-type, since the former exhibits no roots along the $x$-axis.

Second, the standard approach of analyzing the behavior of $\mathbf{K}(x)$ in the vicinity of a root $x = \chi_i$, by using higher-order derivatives $\hat{\varphi}^{(n)}(x)$ to distinguish extrema from flat inflections, does not work in this case. Even the first derivative, $\mathbf{K}'(x) = \operatorname{sgn}\bigl(\hat{\varphi}'(x)\bigr)\,\hat{\varphi}''(x)$,
inherits the unknown sign represented by the function $\operatorname{sgn}\bigl(\hat{\varphi}'(x)\bigr)$.
As a result, only $\mathbf{K}(x)$ itself remains available for the analysis. 

\begin{figure}[t]
\centering
\includegraphics[width=0.8\linewidth]{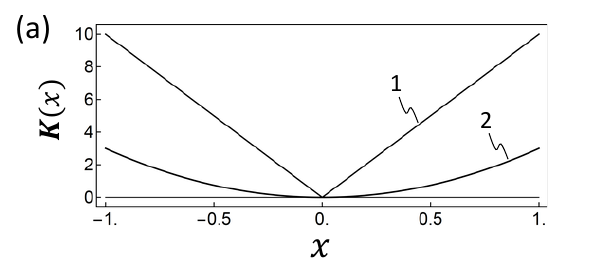}
\caption{Two types of $\mathbf{K}(x)$ behaviour in the vicinity of a root of 
Eq.~(\ref{eq13}) taken from Example 3, for the phase profile see Fig.~\ref{fig:fig3}(b). 
A kink type root (1) along the $x$-axis and a DM type root (2) along the $y$-axis 
are shown. The kink root corresponds to extremum, while the DM root is non-obvious and in this case identified as flat inflection.}
\label{fig:fig7}
\end{figure}

This leads to the third observation: analyzing the behavior of $\mathbf{K}(x)$ in the vicinity of a root $x = \chi_i$ can help distinguish obvious from non-obvious extrema (see Fig.~\ref{fig:fig7}). In such cases, the local profile of $\mathbf{K}(x)$ may exhibit one of several characteristic shapes:

\emph{Kink:} $\mathbf{K}(x) \to 0$ as $c|x-\chi|$, with $c>0$, corresponding to a local phase shape $\hat{\varphi}(x) \approx (x-\chi)^2$ and representing an obvious extremum;

\emph{Cusp:} $\mathbf{K}(x) \to 0$ as $|x-\chi|^{\alpha}$, where $0<\alpha<1$, 
corresponding to $\hat{\varphi}(x) \approx |x-\chi|^{\alpha+1}$, also representing an obvious extremum; 

\emph{Differentiable minimum (DM):} 
$\mathbf{K}(x) \to 0$ as $|x-\chi|^\beta$, where $\beta>1$, corresponding 
to $\hat{\varphi}(x) \approx |x-\chi|^{\beta+1}$, and leading to non-obvious extremum. 

Both the kink $\mathcal{P}_{\mathrm{kink}} \subset \mathcal{P}_{\mathrm{ext}}$ and cusp $\mathcal{P}_{\mathrm{cusp}} \subset \mathcal{P}_{\mathrm{ext}}$ classes yield correct signs for the integration sequence $\{s_{x,i}\}$. The same time for the DM case, no definitive conclusion can be drawn about the root type.
  
Fourth, we address the treatment of non-obvious extrema of the DM type. 
At such points, the standard sign alternation rule can be relaxed: each ambiguous root may take either $s_{x,i}=+1$ or $s_{x,i}=-1$. 
If the sign changes relative to the preceding interval, the point is classified as an extremum; if the sign remains the same, it is classified as a flat inflection. 
For $p$ such ambiguous points in a given slice, this results in $2^p$ possible phase profiles.

To select the physically correct profile, we need knowledge of the absolute phase value at the end of integration path for each slice direction. 
In addition to the initial condition $
\varphi(x_{\min},y_{\min})=\Phi_{00}$,
we must know the absolute phases at the three other domain corners:
\[
\begin{aligned}
\varphi(x_{\min},y_{\max})=\Phi_{01}, \\
\varphi(x_{\max},y_{\max})=\Phi_{11}, \\
\varphi(x_{\max},y_{\min})=\Phi_{10}.
\end{aligned}
\]
These values determine the boundary conditions 
$\varphi(x_{\min},y)$, $\varphi(x_{\max},y)$, $\varphi(x,y_{\min})$, and $\varphi(x,y_{\max})$, which in turn allow selection of the correct profile from the $2^p$ possibilities.

One of the four corner values $\Phi_{ij}$ can always be chosen as the origin of the phase profile and is therefore trivial to define. The remaining three values must be determined experimentally. In other words, for complicated phase profiles susceptible to ambiguous roots of $\mathbf{K}(x)$, the experiment must be designed to provide these $\Phi_{ij}$ or their mutual relations. For example, they may all be set to zero, as in our analytical example; constrained to be equal, as in Example~1; or related by specific conditions, as in Example~7.

In summary, roots of $\mathbf{K}(x)$ may correspond either to extrema, which employ sign alternation, or to flat inflections, which do not. $p$ ambiguous roots yield $2^p$ possible phase profiles, which can be reduced to the true one if three corner phases $\Phi_{ij}$ are known from the experiments. 

\subsection{ Assembling Phase Segments Together}

In traditional approaches, trigonometric inversion yields a number of wrapped phase segments equal to the number of fringes. Increasing the maximal phase amplitude increases the fringe count, and thus the number of segments that must be unwrapped and reassembled.  

Our method avoids fringe-defined segmentation. Instead, phase intervals $s_i$ are determined by the sign of the phase derivative $\sigma_i$, with their boundaries set by the extrema $\chi_i$ and $\gamma_i$ along the $x$- and $y$-axes. The number of that intervals thus reflects the intrinsic functional complexity of the phase rather than its amplitude or fringe density.  

Figure~\ref{fig1} illustrates the difference: six fringes in the traditional method require six wrapped segments, whereas Eq.~(\ref{eq16}) yields only one root, giving two intervals $s_1$ and $s_2$, sufficient for full phase recovery. Even if the phase amplitude were increased to produce 100 fringes, the same two intervals would be used.  

This extrema-based segmentation fundamentally distinguishes our approach from traditional fringe-based methods.

\subsection{ Multiple-Beam Interference}

To illustrate the adaptability of our first-principles phase-retrieval approach, we present, for reference, the interferogram function and corresponding phase-retrieval equation for multiple-beam interference, applicable to thin-film configurations.

The interferogram function $F$ for multiple-beam interference is given by
\begin{equation}
F = \frac{\kappa + \beta - 2kG}{-\kappa - \beta + 2\beta G},
\label{eq24}
\end{equation}
where $\beta = 2r_1 r_2$ and $\kappa = 1 + r_1^2 r_2^2$ are coefficients 
derived from the Fresnel reflection coefficients, as described in \cite{Sujanani1991,Gokhale2004}, 
and where the gray function $G$ has a uniform intensity envelop. The Fresnel coefficients $r_1$ and $r_2$ 
for this case are defined as
\[
r_1 = \frac{n_0 - n_1}{n_0 + n_1}, \quad r_2 = \frac{n_1 - n_2}{n_1 + n_2},
\]
where $n_0$ is the refractive index of the medium from which the incident beam 
originates, $n_1$ is the refractive index of the thin film (i.e., the object beam material), 
and $n_2$ is the refractive index of the medium into which the transmitted beam exits. 
These notations follow the conventions established in \cite{Heavens1965}. 
In this formulation, the gray function $G$ lies in the range $G \in [0, 1]$, 
in contrast to the idealized two-beam interference case, where $G \in [0, 2]$.

The corresponding phase-retrieving differential equations for multiple-beam interference read
\begin{equation}
(\varphi'_x)^2 = \frac{(F'_x)^2}{4(1 - F^2)}, \
(\varphi'_y)^2 = \frac{(F'_y)^2}{4(1 - F^2)},
\label{eq25}
\end{equation}
where the factor $4$ arises due to the double passage of the object beam through the thin film, 
resulting in the $\cos(2\varphi)$ term in Eq.~(\ref{eq4}). 
Here, $\varphi$ corresponds to the spatial phase difference introduced by the thin film with refractive index $n_1$.

While the phase-retrieving differential equation differs from Eq.~(\ref{eq6}) 
only by a constant factor, the interferogram function itself is substantially 
different, both in shape and in the range of the gray function intensity envelop.

\subsection{ Towards Practical Interferograms}

The proposed phase-retrieval method is demonstrated under conditions where the integrand $\mathbf{K}$ in Eq.~(\ref{eq16}) depends solely on the phase. In this case, its roots are determined exclusively by the phase, allowing accurate phase reconstruction through integration. This behavior is ensured by the specific structure of Eq.~(\ref{eq16}), which requires the interferogram function $F$ to exhibit a uniform intensity envelope confined to the interval $[-1, 1]$.

The simplest way to obtain such interferogram function $F$ is to begin with a function $G$ that directly inherits the uniform intensity envelope from the original interferogram. This approach can be adopted in our demonstrations. By setting $A = B = 1$, we can construct an interferogram with a uniform envelope, resulting in $G \in [0, 2]$. From this, the corresponding $F$ can be derived by using the envelope normalization 
\begin{equation}
F=\frac{2G - (G_{max} + G_{min})}{G_{max} - G_{min}},
\label{eq26}
\end{equation}
instead of Eq.~(\ref{eq7}), with $G_{min}=0$ and $G_{max}=2$, respectively.

In practice, real interferograms rarely exhibit the uniform intensity envelope. Instead, the observed fringes typically follow a complex, non-uniform spatial intensity modulation, representing $\tilde{G}$ with a non-uniform intensity envelop. In the case of idealized two-beam interference, this modulation is described by functions $A(x)$ and $B(x)$ in the form
$\tilde{G}(x) = A(x) + B(x)\cos\varphi(x)$. 
These functions, which are generally unknown, independently modulate both the background and the contrast of the fringes.

Because $A(x)$ and $B(x)$ are not known a priori, generalizing the proposed method directly to such cases is not feasible. However, a practical alternative is to introduce a preprocessing step that flattens the non-uniform envelope of the original interferogram $\tilde{G}(x)$. This flattening transforms $\tilde{G}(x)$ into an effective $G(x)$ whose maxima and minima conform to the uniform envelope condition. As a result, the transformed interferogram becomes suitable for direct application of our method.

\textit{Illumination background and amplitude modulation:} The concept of flattening a general non-uniform $\tilde{G}$ is outlined in \cite{Sujanani1991,Gokhale2004}, representing a generalized version of Eq.~(\ref{eq26}). In brief, it involves constructing two envelope functions, $\tilde{G}_{\downarrow}(x)$ and $\tilde{G}_{\uparrow}(x)$, which bound the fringe intensity from below and above, respectively. Normalizing $\tilde{G}$ between these envelopes produces the flattened version of $G \in [0, 1]$, yielding a corresponding $F \in [-1, 1]$ compatible with our method \cite{Berejnov2010}. In such cases, both $G$ and $F$ can typically be approximated by smooth functions.

This approximation, however, introduces additional numerical subtleties. Even at rational coordinates, the integrand $\mathbf{K}$ may produce indeterminate forms of type $0/0$. These points must be identified and excluded before numerical integration. At such points, the integrand must be replaced using the square root of the sign-specific second derivative of $F$, according to the extremum type, in accordance with l’Hospital’s rule.

\textit{Noise in fringe intensities:} In practice, interferograms are subject to noise at the pixel level, resulting in a noisy gray-level function $\tilde{G}$. Flattening $\tilde{G}$ and approximating it with a smooth function may still yield an acceptable $F \in [-1, 1]$, however, this alone is insufficient. Since the integrand $\mathbf{K}$ depends on the derivative $F'(x)$, high-frequency noise in $F$ can introduce spurious components into $\mathbf{K}$. These artefacts may obscure the true roots, creating numerous pseudo-roots and generating artificial integration segments $s_i$. Convolving the integrand $\mathbf{K}$ with a Gaussian kernel may improves its smoothness and helps to retain only the true roots. While effective in some examples, this smoothing technique may not be universally applicable.

\subsection{ Comparison with Other Methods}

Phase-recovery techniques in interferometry cover a broad spectrum \cite{Judge1994, Baldi2001}, but the range narrows considerably for single-interferogram analysis, which can be divided into two categories.

The first category comprises numerical algorithms that directly invert the trigonometric phase–intensity relation. Their main difficulty lies in determining the sign of the phase derivative across the folded fringe pattern, which is essential for correct unfolding. Because the interferogram fringe intensity alone does not reveal whether the phase is increasing or decreasing, identifying sign changes is nontrivial and typically relies on additional heuristics \cite{Okada2007}. As a result, the conversion from folded to wrapped phase often depends on simplified assumptions about phase gradient sign transitions. In this category, the subsequent unwrapping step is generally not the limiting factor.

The second category includes the Fourier Transform (FT) method, originally proposed in \cite{Takeda1982} and applied for Fizeau-type interferograms. By introducing a spatial carrier frequency along one axis, the method linearly modulates the phase to prevent folding, producing a wrapped phase that can be readily unwrapped along that modulation direction. The sign of the phase derivative orthogonal to the carrier is fixed by constant boundary conditions, enabling complete phase reconstruction. However, the FT method is not applicable to Newton-type interferograms, which lack such modulation and thus retain sign ambiguity.   
The limitation of the FT method is the need to heuristically define phase boundary conditions in the direction orthogonal to unwrapping.

Both classes suffer from an inherent phase-reconstruction degeneracy, a fundamental limitation of single-interferogram analysis, in which the recovered phase may appear inverted in sign or orientation. Neither class can be considered analytical, 
as both rely on processing local fringe segments. The first class requires heuristic knowledge of the phase-gradient signs, while the second is restricted to Fizeau-type interferograms. In contrast, our method automatically determines the phase-gradient signs and is equally applicable to both Newton- and Fizeau-type interferograms.

In this context, it is worth mentioning an alternative approach based on 
the Transport of Intensity Equation (TIE), introduced  in \cite{Teague1983} and 
later applied in interferometry simulation  \cite{Pandey2016}. The TIE method involves solving a three-dimensional partial differential equation to recover a continuous global phase field. While it indeed yields a continuous solution, it requires multiple interferograms sequentially recorded at different focal planes. This requirement places the method outside the scope of single-interferogram analysis and thus beyond the focus of the present work.

\section{ Conclusion}

The method introduced in this paper represents a fundamentally new class 
of continuous phase-retrieval techniques for single interferograms. Unlike the traditional 
approaches, it operates with an ordinary differential equation that relates the 
interferogram function to the underlying phase. This enables the continuous recovery 
of a phase profile between any two points on the interferogram, 
thereby eliminating the need for phase unfolding and unwrapping. As a result, the method avoids complications inherent to heuristic conditions of the phase 
itself and its derivative, fringe-wise operation, and  underestimation of the phase at fringe discontinuities while reconstructing.

Our method accommodates arbitrary phase boundary conditions, which can either be predefined or extracted from the interferogram during the reconstruction process. It is formulated within the framework of explicit calculus, making it, to the best of our knowledge, the first analytical tool for phase retrieval from a single interferogram, provided certain idealizations and well-behaved conditions are satisfied.

The method is equally applicable to both Newton-type and Fizeau-type interferograms. In the Fizeau configuration, an additional simplification may arise from the elimination or reduction of phase extrema along the modulation direction.

Several important considerations must be addressed when applying this method:

\begin{enumerate}
\item \textit{Degeneracy of global sign:} Like all single-interferogram techniques, our method is subject to phase-reconstruction degeneracy with respect to the global sign of the reconstructed phase. To resolve this, the reader may need to adjust the initial signs, such as $\sigma_{x,1}$ or $\sigma_{y,1}$, to align the reconstructed phase with known experimental conditions.

\item \textit{Root identification:} The method requires solving an algebraic equation $\mathbf{K}^2=0$
for the integrand square  to determine roots $\chi_i$ and $\gamma_i$ along the respective $x$- or $y$-axes.  Roots corresponding to flat inflection points, where the phase derivative vanishes without changing sign, must be excluded, 
as they do not sepatate the intervals of phase change. The remaining 
extrema define the integration segments $s_i$ within which the phase derivative has a constant sign.  
    
\item \textit{Uniformity of fringe intensity:} The method is designed to operate on an interferogram function $F$ whose fully developed fringes have a constant amplitude, bounded within the interval $[-1, 1]$, representing purely phase modulation. If the fringe intensities of the initial interferogram are modulated by background illumination or contrast variations, preprocessing is required to determine a uniform intensity envelope and apply it to the interferogram.

\item \textit{Conditions of interference:} The mathematical form of the interferogram function (e.g., Eqs.~(\ref{eq3}) and (\ref{eq24})) and the corresponding differential equations (Eqs.~(\ref{eq6}) and (\ref{eq25})) are sensitive to the interferometric configuration and the specific interference conditions of the experiment. Accurate formulation requires incorporating these experimental parameters.
\end{enumerate}

Overall, we anticipate that this new method can support a wide range of applications that inherently rely on a single interferogram as the primary data source. For Newton-type interferograms, representative potential examples could include the reconstruction of profiles of moving liquid menisci~\cite{DellAversana1997} and the static curvature of thin liquid films~\cite{Gokhale2004,Berejnov2010}. In the case of Fizeau-type interferograms, the method may help the determination of spatial distributions of concentration~\cite{Mialdun2011} and temperature~\cite{Bratukhin2005}.   

In conclusion, this analytical method introduces a new perspective for phase retrieval 
from a single interferogram and opens opportunities for its application across a broad 
range of optical, fluidic, and materials research domains.

\newpage
\section{Figures and Captions}

\end{document}